\documentclass{article}
\usepackage{amsmath,amssymb}
\usepackage{graphicx}
\usepackage[left=2.54cm, right=2.54cm, top=2.54cm,bottom=2.54cm]{geometry}
\usepackage{subfigure}
\usepackage{color}
\usepackage{wrapfig}
\usepackage{wasysym}
\usepackage[all]{nowidow}
\usepackage[normalem]{ulem}
\usepackage{soul}
\usepackage{lipsum}
\usepackage{booktabs,xltabular}
\usepackage{appendix}
\usepackage{enumitem}

\usepackage{sectsty}

\sectionfont{\fontsize{12}{15}\selectfont}
\subsectionfont{\fontsize{10}{12}\selectfont}

\usepackage[american]{babel}
\usepackage{csquotes}
\usepackage[style=apa, backend=biber]{biblatex}
\DeclareLanguageMapping{american}{american-apa}
\DeclareDelimFormat{nameyeardelim}{\space}

\setlist[enumerate]{itemsep=0mm}

\begin{document}

\begin{center}
{\bf \Large Extending the Use of Information Theory in Segregation Analyses to Construct Comprehensive Models of Segregation} 
\end{center}

\begin{center}
    {Boris Barron$^{1,2}$, Yunus A. Kinkhabwala$^3$, Chris Hess$^{2,4}$, Matthew Hall$^{2}$, Itai Cohen$^{1}$, Tom\'{a}s  A. Arias$^{1}$}
\end{center}

\begin{small}
\begin{center}{
    $^1$Department of Physics, Cornell University, Ithaca, NY 14853, USA. \\
    $^2$Department of Policy Analysis, Cornell University, Ithaca, NY 14853, USA. \\
    $^3$Department of Applied and Engineering Physics, Cornell University, Ithaca, NY 14853, USA. \\
    $^4$Department of Sociology and Criminal Justice, Kennesaw State University, Kennesaw, GA 30144, USA. \\
    }
\end{center}
\end{small}

\begin{center}
\section*{Abstract}
\end{center}

The traditional approach to the quantitative study of segregation is to employ indices that are selected by ``desirable properties''. Here, we detail how information theory underpins entropy-based indices and demonstrate how desirable properties can be used to systematically construct models of segregation. The resulting models capture all indices which satisfy the selected properties and provide new insights, such as how the entropy index presumes a particular form of intergroup interactions and how the dissimilarity index depends on the regional composition. Additionally, our approach reveals that functions, rather than indices, tend to be necessary mathematical tools for a comprehensive quantification of segregation. We then proceed with exploratory considerations of two-group residential segregation, finding striking similarities in major U.S. cities, subtle segregation patterns that correlate with minority group diversity, and substantive reductions in segregation that may be overlooked with traditional approaches. Finally, we explore the promise of our approach for segregation forecasting.

\vspace{0.4cm}
\section{Introduction}
Residential segregation is widely acknowledged as being a central axis of stratification, particularly in the U.S., where it has famously been described as being the ``structural linchpin" of inequality (\cite{pettigrew1979racial}; \cite{massey2016residential}; \cite{charles2003dynamics}). Residential segregation correlates with the unequal racial exposure to poverty (\cite{logan2011separate}), crime (\cite{akins2007racial}), pollution (\cite{crowder2010interneighborhood}) and is fundamental to unequal outcomes in areas such as earnings (\cite{thomas2015race}) and health (\cite{kershaw2015racial}). Residential segregation has roots in historical policy (\cite{nodjimbadem2017racial}) and is one of the most prominent manifestations of racism~(\cite{popescu2018racial}).  

The quantitative study of residential segregation is typically performed by analyzing how populations in a region distribute into small-scale neighborhoods. These small-scale units are usually delineated as census tracts or census blocks, while the region is a county or metropolitan area. `Segregation' is then the seemingly persistent disassociation of various groups, and the study of segregation centers around understanding its causes, histories, variability, and implications. Due to the sheer complexity present in human systems, segregation is a multi-faceted phenomenon that cannot be easily understood: regions are distinct and individuals idiosyncratic, and it is difficult to develop even coarse-grained descriptions of intergroup interactions.

Traditionally, the quantification of segregation has relied on \emph{segregation indices}, which provide values for comparison across space and time. However, despite the plethora of indices that have been introduced, finding the `best' index, or even the `best' set of indices, remains an unsolved problem (\cite{allen2007should}; \cite{frankel2011measuring}). Segregation analyses have thus devoted considerable attention to identifying properties of segregation measures that would be desirable, the so-called \emph{desirable properties}. Study of properties and various indices leads to the conclusion that every index suffers from some flaw or drawback (\cite{frankel2011measuring}). Given such difficulties, it is then unsurprising that indices which have found widespread use are those with exceptional conceptual simplicity rather than rigor or utility: there is no perfect index.

This manuscript details how to employ \emph{desirable properties} not as benchmarks of segregation measures but as defining characteristics in obtaining a quantification of segregation. As we demonstrate, the resulting quantification can then be used in tandem with traditional indices and underpin models that create powerful index counterfactuals. Furthermore, in practice, our approach identifies trends that can be obscured with traditional index analyses and opens new avenues in the study of segregation, including clustering regions by their detailed segregation trends and incorporation of segregation into population forecasting.

\section{Data}
This manuscript uses decennial census data for metropolitan statistical areas at the block-group level for the 1990-2020 decades (\cite{Mason2022nhgis}). Moreover, `White' and `Black' are defined as non-Hispanic White and non-Hispanic Black, respectively, while `Non-White' encompasses all groups other than White, and `Other' indicates individuals that are not White, Black, or Hispanic.

\section{Notation}
We employ the notation of Reardon and Firebaugh (2002\nocite{reardon2002measures}): $t$ denotes the neighborhood size (population) and $\pi$ denotes composition (proportion); subscript $i$ indicates the organizational unit, what we call neighborhoods, and the overall system in which these neighborhoods reside we call a region. Furthermore, subgroups (\textit{e.g.,} racial group) are indicated by either $m$ or $n$. More precisely, 
\begin{equation*}
\begin{aligned}
   T &  =  \text{total number of individuals in the region}\\
   T_{m} & =  \text{total number of individuals of subgroup $m$ in the region}\\
   \pi_m & = \text{composition of subgroup  $m$ in the region, $\pi_m = T_m/T$}\\
   t_i & =  \text{number of individuals in neighborhood $i$: $ \textstyle \sum_i $ $t_i =T$}\\
   t_{im} & =  \text{number of individuals of subgroup $m$ in neighborhood $i$: $\textstyle \sum_i $ $t_{im} = T_m$} \\
  \pi_{im} & = \text{composition of subgroup $m$ in neighborhood $i$, $\pi_{im} = t_{im}/t_i$}.\\
\end{aligned}
\end{equation*}
Moreover, there is often a need to succinctly write the subgroup makeup of a neighborhood (\emph{e.g.,} occupancy of 400 White, 50 Black, 200 Hispanic, ...), the notation we shall use for this is
\begin{equation*}
\begin{aligned}
   \{t_i\} \equiv  & (t_{i1}, t_{i2}, t_{i3}, ...) = \text{ the occupancy of neighborhood $i$} \\
    \{\pi_i\} \equiv & (\pi_{i1}, \pi_{i2}, \pi_{i3}, ...) =  \text{ the composition of neighborhood $i$, $\{ \pi _i\}   = \Big(\frac{t_{i1}}{t_i}, \frac{t_{i2}}{t_i}, \frac{t_{i3}}{t_i}, ...\Big)$} . \\
\end{aligned}
\end{equation*}
When the subscript is dropped in `$\{ \}$' (\emph{e.g.,} $\{t\}$) then it will still refer to circumstances at the neighborhood-level but the precise neighborhood is then either irrelevant or unambiguous.

\section{Common Segregation Indices} \label{sec:common-inds}

The commonly used ``segregation dimensions'' are those of \emph{evenness}, how uniformly racial subgroups are distributed throughout a region, and \emph{exposure}, how frequently individuals from various subgroups encounter each other (\cite{massey1988dimensions}). For \emph{evenness}, the most common measure is the dissimilarity index. For the binary scenario in which individuals are categorized into only two groups (\emph{e.g}., White/non-White), the dissimilarity index can be written
\begin{equation}
\begin{aligned}
    D = \frac{1}{2T} \frac{1}{\pi_m(1-\pi_m)} \sum_i  t_i |\pi_{im} - \pi_{m}|.
\end{aligned} \label{dissimilarity2}
\end{equation}
The dissimilarity index measures the fraction of a subgroup that would need to be moved for every populated neighborhood to have the same binary composition. For multigroup scenarios, or where additive decomposability is important, another commonly used measure of evenness is the entropy index,
\begin{equation}
\begin{aligned}
    S = \frac{1}{TE}\sum_m \sum_i t_i \pi_{im} \text{ln} \Big(\frac{\pi_{im}}{\pi_{m}} \Big), \text{ with } E = -\sum_m \pi_{m} \ln (\pi_{m}).
\end{aligned} \label{entropy_index}
\end{equation}

For \emph{exposure}, the most common measure is the interaction index, 
\begin{equation}
  _n P^*_m = \sum_i \frac{t_{in}}{T_n} \frac{t_{im}}{t_i},
\end{equation}
which measures what fraction of the time an individual of subgroup $n$ is expected to encounter an individual from subgroup $m$ as opposed to a member of a different group. As the index is generally not symmetric between $m$ and $n$, we follow the standard convention of taking the minority group perspective; specifically, we measure how often a member of the minority group is expected to encounter a member of the White group. Further, if $m$ and $n$ constitute the same group, the measure is instead known as the isolation index
\begin{equation}
 {}_m P_m^* = \sum_i \frac{t_{im}}{T_m} \frac{t_{im}}{t_i}.
\end{equation} 

We focus on the above indices as they are among the most common and can be used to explore various properties of the methods we develop. In particular, the binary dissimilarity index and the interaction index are useful as they are on opposite extremes in terms of their dependence on the regional composition: the interaction index has a strong correlation with the overall composition in a region, while the dissimilarity index is typically considered to, at most, have a weak dependence. The reason we explore the entropy index is because it is among the most rigorous (\cite{reardon2002measures}) and is one of the most closely related indices to our approach. 

\section{Limitations of Traditional Segregation Approaches}

\subsection{Discordance of Indices and Properties} \label{sec:discord}

A fundamental issue with segregation indices is the disjointed nature in which they connect to \emph{desirable properties}. Traditionally, measures are chosen to satisfy a set of properties (\emph{e.g.,} size-invariance or organizational equivalence); however, unless these properties result in a unique mapping of data to an index, every index must necessarily introduce additional `undesirable' assumptions. Our proposal to resolve this is then the following: (1) simplify data precisely as much as a choice of properties allows and (2) require that a mapping of this `simplified data' to a measure of segregation be unique. In this section we will demonstrate how properties can be used to simplify data directly, and the following sections detail how the entirety of the resulting `simplified data' can be utilized. 

As an illustration, suppose we have standard aspatial demographic data for a region (\emph{e.g.}, neighborhood \emph{A} has 400 White, 50 Black and 200 Hispanic individuals; neighborhood \emph{B} has...; \emph{etc.}) and we are trying to quantify segregation. To start, suppose a selected property, meaning a property we choose to be satisfied by our measures of segregation, is \emph{neighborhood ordering irrelevance}, such that the labelling of neighborhood \emph{A} and neighborhood \emph{B} is not meaningful. In this common situation, the data can be simplified to a counting of how many neighborhoods have a particular occupancy $\{t\}$, 
\begin{equation}
\begin{aligned}
    G\{t\} = \sum_{i} \delta(\{t\},\{t_i\}),
\end{aligned}  \label{eq:aspatial0}
\end{equation}
where $\delta(\{t\},\{t_i\}) = 1$ if $\{t_i\} = \{t\}$, and $\delta(\{t\},\{t_i\}) = 0$ otherwise. We emphasize that conceptually what this represents is quite simple: a value of $G(400, 50, 200) = 5$ means that there are exactly 5 neighborhoods in the region that have an occupancy consisting of 400 White, 50 Black, and 200 Hispanic individuals. The quantity $\delta((400, 50, 200),\{t_i\})$ is then simply used to check if neighborhood \textit{i} is (400, 50, 200), returning 1 only if it is indeed (400, 50, 200) and returning 0 otherwise. Summing over all neighborhoods then becomes a counting of how many neighborhoods are (400, 50, 200) in the region. Of course, \textit{any} occupancy can be selected, not just (400, 50, 200), and so a quantification of $G\{t\}$, for all possible $\{t\}$, then fully represents the initial data set with only neighborhood ordering removed. Hence, any measure (\textit{e.g.,} segregation index) that is not affected by neighborhood ordering could just as well be calculated from $G\{t\}$ as it could from the initial data set. Finally, we emphasize that there are many initial data sets that could lead to the same `simplified data', represented here by $G\{t\}$; however, for the $G\{t\}$ function to be the same for different initial data sets, it must be that these data sets differed \emph{only} in the ordering of their neighborhoods which, in this case, we explicitly intended to make irrelevant. 

Continuing to simplifying the data, we note that the ``size invariance'' \emph{desirable property} states that segregation is unchanged if all neighborhoods in a region are duplicated and the resulting system is
treated as a single large region (\cite{james1985measures}). To incorporate this property, the counting function $G\{t\}$ must be normalized by the number of neighborhoods in the region $N_R$. This leads to
\begin{equation}
\begin{aligned}
    P\{t\} = \frac{1}{N_R} \sum_{i} \delta(\{t\},\{t_i\}),
\end{aligned}  \label{eq:aspatial}
\end{equation}
which can be interpreted as the probability of randomly picking a neighborhood in a region and finding it to possess an occupancy $\{t\}$.

A more subtle \emph{desirable property} is ``organizational equivalence'', which states that neighborhoods with the same level of segregation (same subgroup compositions) can be combined with segregation remaining unchanged (\cite{james1985measures}). This implies, for example, that a region having two neighborhoods that are each 650 in size with occupancies (400, 50, 200) could be treated as having, instead, a single neighborhood that is 1300 in size with an occupancy of (800, 100, 400). Hence, the primary quantity of interest becomes neighborhood compositions, instead of occupancies, and the contribution of neighborhoods becomes size-weighted. Such a data simplification can be computed from $P\{t\}$ in equation (\ref{eq:aspatial}) by performing a size-weighted sum of occupancies resulting in the same composition or, equivalently, computed directly from the initial data set by using
\begin{equation}
\begin{aligned}
    P\{\pi\} = \frac{1}{T} \sum_{i} t_i \delta(\{\pi\},\{\pi_i\}).
\end{aligned} \label{eq:org_inv0}
\end{equation}
Note that the normalization has been changed from the number of neighborhoods in the region $N_R$ to the total number of individuals $T$ so that the size-invariance property remains satisfied. This version of `simplified data', the $P\{\pi\}$ function, we refer to as the \textit{compositional distribution}. This distribution represents the probability of randomly picking an individual from a region and finding this individual to be in a neighborhood with composition $\{\pi\}$. 

If there is a desire to consider binary comparisons, such as White/non-White or White/Black, the other groups can be omitted (\textit{e.g.,} ignoring all individuals other than White and Black for White/Black segregation), leading to a single-variable compositional distribution
\begin{equation}
\begin{aligned}
    P(\pi) = \frac{1}{T} \sum_{i} t_i \delta(\pi,\pi_{im}),
\end{aligned} \label{eq:org_inv}
\end{equation}
where $\pi$ indicates the composition of a group of interest (\emph{e.g.}, White), and $\pi_{im}$ this group's composition in neighborhood $i$.

Considering that typical \emph{desirable properties} lead to data simplification to the forms found in equations \ref{eq:aspatial0}, \ref{eq:aspatial}, \ref{eq:org_inv0} and \ref{eq:org_inv}, it is evident that the proper mathematical objects for characterizing segregation tend to be \emph{functions}, rather than \emph{indices}. This is perhaps not surprising, the reduction of demographic data to an index constitutes such a substantial loss of information that this simplification could not reasonably be achieved with any widely-accepted set of properties. Nevertheless, such arduous simplification to an index has been attempted~(\cite{hutchens2004one}). The resulting ``square-root index'', however, has remained in relative obscurity for two reasons: its conceptual complexity and use of uncommon desirable properties. By not requiring that the resulting measures be an index, a reasonable set of properties can be selected that is not dictated by an inflexible level of simplicity.

Further, we emphasize that appropriate simplification of demographic data does indeed preserve all relevant information consistent with the employed properties. To illustrate this, Table~1 shows how \emph{all} binary indices, which satisfy the aforementioned properties, can be computed directly from the compositional distribution which can be accomplished because these indices can be represented generally as a size-weighted average of a function depending on neighborhood compositions (first line of Table 1). 

Finally, we note that there is, of course, substantial freedom when it comes to choosing \emph{desirable properties}, and many works have focused precisely on determining which are relevant (\cite{allen2007should}; \cite{frankel2011measuring}). In this work, we predominantly consider the aforementioned properties that can be summarized as being ``aspatial" and ``scale invariant", as these are properties that are found in nearly all commonly-used segregation indices (\cite{frankel2011measuring}).

\subsection{Controversial Properties: Compositional Invariance} \label{sec:comp-inv}

We have demonstrated how to incorporate a number of common \emph{desirable properties}, but sometimes a property itself is controversial or unclear. A notable example of such a property is \emph{compositional invariance}. Broadly, compositional invariance intends to remove the aspect of segregation resulting \emph{only} from the regional composition, thus making it possible to compare regions whose compositions differ (\textit{e.g.,} comparing segregation in a city that is 20\% White overall to a city that is 60\% White). Not only are there practical difficulties in how to mathematically define such a property (\cite{kalter2000measuring}; \cite{coleman1982achievement}; \cite{reardon2004measures}), the concept itself is contentious because segregation, as `experienced' by an individual, is dependent on the regional composition (\emph{e.g.,} interaction index). Compositional invariance is then problematic from both practical and conceptual standpoints. 

In that case, why bother with such a controversial property? The answer is that, when comparing segregation between different regions, the issue is unavoidable. Regional compositions do differ, and so to ignore differences in regional compositions is to imply that measures of segregation are independent of such differences, an assumption which is no less controversial. The traditional resolution is to consider a middle ground, a partition of a segregation index into a sum of ``marginal'' and ``structural'' components, with the marginal components accounting for circumstantial factors, such as neighborhood size and regional composition, and structural components accounting for the ``pure segregation'' (\cite{elbers2021method}). Then, \textit{if} the marginal components can be properly accounted for, structural components can be isolated and segregation for regions can be compared in a compositionally invariant manner.

Our proposal, that we elucidate in the proceeding sections, is notably distinct from traditional approaches in that we do not perform such a partitioning between marginal and structural components. Instead, we will use information theory to determine the least-biased \emph{transformation} that a change in the regional composition would create in our simplified data, allowing us to modify the compositional distribution directly. We will then consider \emph{structural segregation} to be the aspect of our simplified data that is unchanged for different regional compositions. Throughout this manuscript, we shall refer to the `compositionally invariant aspect of the compositional distribution' as \emph{compositional behavior}.

We develop the above ideas in the next two sections as follows. First, we use information theory to construct a model of segregation from only marginal information (neighborhood size and regional composition). A model which allows us to demonstrate that traditional entropic indices incorporate, in a very simplistic sense, a notion of compositional invariance based upon subgroup-dependent interactions. We then extend this procedure to the compositional distribution and demonstrate how a much more detailed accounting of subgroup-dependent interactions can be incorporated into a compositionally invariant `segregation function'.

\section{Entropic Indices and Non-Interacting Systems}

\subsection{Entropic Indices Through the Lens of Information Theory} \label{sec:ent-ind-info}

Given our intention to make maximal use of information contained in data satisfying a set of properties, the appropriate branch of mathematics for developing improved measures of segregation is Information Theory. This field has found rising use in demographic analysis and, as we summarize here, can be used to readily derive the entropic indices. In information theory, the quantification of `uncertainty' (which is the amount of additional information needed to specify an outcome) is accomplished with \emph{entropy},
\begin{equation}
\begin{aligned}
    S\{P\} = -\sum_k P_k \ln P_k,
\end{aligned}
\end{equation}
where $\{P\}$ indicates the collection of $k$ probability outcomes, $\{P\} = \{P_1, P_2, P_3, ...\}$. (See Appendix \ref{info theory} for background information.) Note that $P_k$ is always taken to be a true, properly normalized, probability distribution such that $\sum_k P_k = 1$.  As well, entropy  remains unaffected when including `potential' outcomes that never actually occur, $P_k \ln P_k = 0$ if $P_k = 0$, and, as entropy is the quantification of uncertainty, it is appropriately strictly non-negative.

In terms of demographic analysis, the multigroup entropy index (\cite{reardon2002measures}) is computed as
\begin{equation}
\begin{aligned}
    S = \frac{1}{TE}\sum_m \sum_i t_i \pi_{im} \text{ln} \Big(\frac{\pi_{im}}{\pi_{m}} \Big), \text{ with } E = -\sum_m \pi_{m} \ln (\pi_{m}).
\end{aligned} \label{entropy_index2}
\end{equation}
This measure has its minimum possible value of zero when $\pi_{im} = \pi_m$ for all neighborhoods, corresponding to every neighborhood having the same composition as the region overall, \textit{i.e.,} complete integration. Its maximum value is one, which occurs when $\pi_{im} =0$ or $1$ for all neighborhoods, corresponding to every neighborhood consisting of entirely a single subgroup, \emph{i.e.}, complete segregation. More recently, the contribution to this index from a single person in a given neighborhood $i$ has been referred to as that neighborhood's divergence index (\cite{roberto2015divergence}),
\begin{equation}
\begin{aligned}
   D_i = \sum_m \pi_{im} \ln \frac{\pi_{im}}{\pi_m}. 
\end{aligned}
\end{equation}
Both of these measures are closely related to the original entropy index as presented by Theil and Finizza (1971\nocite{theil1971note}) which omits the normalization by \textit{E} in equation \ref{entropy_index2}, is referred to as the \emph{M} index, and has been useful in its own right (\cite{mora2011entropy}; \cite{elbers2021method}). 

Because the entropy and \textit{M} indices are aspatial and scale-invariant, they can be calculated from the compositional distribution which is the probability of randomly picking an individual in a region and finding the individual to be in a neighborhood with composition $\{\pi\}$. Correspondingly, to derive the entropic indices, the idea is to randomly choose an individual and quantify the uncertainty as to which subgroup this individual belongs given (1) only regional knowledge and (2) neighborhood-level knowledge of compositions. 

Mathematically, consider the uncertainty of finding an individual of subgroup $m$ (\emph{i.e.}, White, Black, \emph{etc.}) in a region of known compositional makeup (\emph{i.e.}, 40\% White, 40\% Black, \emph{etc}.), but \emph{without} knowledge of neighborhood-level compositions. The probability of picking an individual of group $m$ is then $\pi_m$ and the entropy characterizing the aforementioned uncertainty is
\begin{equation}
\begin{aligned}
    S(m|\pi_m) = -\sum_m \pi_m \ln \pi_m.
\end{aligned}
\end{equation}
Next, suppose that we do have knowledge of the neighborhood composition from which the individual is drawn. With this new knowledge, the probability of picking an individual of group $m$ is characterized by the neighborhood-level composition $\pi_{im}$ and the expected (average) uncertainty for the entire regional population becomes
\begin{equation}
\begin{aligned}
    S(m|\pi_{im}) = E\Big[-\sum_m \pi_{im }\ln \pi_{im} \Big] = \frac{1}{T} \sum_i t_i \sum_m \pi_{im }\ln \pi_{im}, 
\end{aligned}
\end{equation}
where $t_i$ is the population of neighborhood $i$, and $T$ is the total regional population. The `reduction in uncertainty' between regional and neighborhood level information is thus
\begin{equation}
\begin{aligned}
    M=S(m|\pi_m) - S(m|\pi_{im}) = E\Big[-\sum_m \pi_{m }\ln \pi_{m}+\sum_m \pi_{im }\ln \pi_{im} \Big]  = \frac{1}{T} \sum_i t_i \sum_m  \pi_{im} \ln \Big(\frac{\pi_{im}}{\pi_m}\Big),
\end{aligned}
\end{equation}
which is precisely the \emph{M} index of Theil and Finizza (1971\nocite{theil1971note}). Note that the \emph{M} index can never be negative because the uncertainty of an individual's subgroup will always be lower provided more detailed information. Finally, normalizing the \emph{M} index by its maximum value, $S(m|\pi_m)$, leads to the appropriately named \emph{entropy index} of Reardon and Firebaugh (2002) from equation~\ref{entropy_index2}. The entropy index has a value of zero when neighborhood-level data is no more informative than regional data, indicating absence of segregation, and has a value of one when neighborhood-level data leaves no remaining uncertainty as to group identity, indicating complete separation of the subgroups. Consequently, we see that information theory is intimately connected to common sociological measures and it is then natural to expect that more sophisticated information theory techniques will result in more powerful tools. 

\subsection{Relation of Entropic Indices to a Non-Interacting Model of Segregation} \label{sec:logliklihood}
Beyond the construction of segregation measures, information theory can be applied much more powerfully to construct detailed models of the underlying joint probability distributions describing the possible neighborhood compositions (or occupancies) in a region. Such a joint probability distribution would indicate, for example, how likely a neighborhood of size 650 is to consist of precisely 400 White, 50 Black and 200 Hispanic individuals based on a certain state of knowledge. In this section, we demonstrate that the construction of these models elucidates the assumptions underlying traditional entropic indices and, in the following section, we describe how to make models that are systematically consistent with an employed set of properties.

The information theory concept that allows construction of the underlying joint probability models is the \emph{principle of maximum entropy} (Appendix \ref{ref:PME}).  We will begin with construction from the most basic information: the regional composition and neighborhood sizes. We will then use the \emph{observed} neighborhood compositions to quantify the accuracy of the resulting model using cross-entropy (Appendix \ref{CE}), which then gives a measure of how well knowledge of the regional composition anticipates the actual neighborhood-level observations.

Specifically, knowledge of the regional composition and size of a neighborhood leads to a unique maximum entropy occupancy model (Appendix \ref{non-int MaxEnt}) given by
\begin{equation} \label{eq:multinomial}
\begin{aligned}
    P\{t\}= \frac{t_i!}{\prod_m t_{im}!} \prod_m \pi_m^{t_{im}},
\end{aligned}
\end{equation}
 which represents precisely the well-known multinomial distribution. For this distribution, the rightmost product gives the probability of independently choosing individuals from each subgroup $m$ a number of times equal to $t_{im}$, given that $t_i$ individuals are chosen in total. Meanwhile, the combinatorial prefactor counts the number of various configurations (\emph{i.e.}, sequences: first person is \emph{White}, second person is \emph{Black}, \emph{etc.}) which lead to the same neighborhood occupancy. The multinomial distribution is thus what would be expected when selecting $t_i$ individuals for a neighborhood completely at random from a region with a composition determined by the $\pi_m$, and is what should be expected if subgroup membership had no influence on how individuals are distributed in neighborhoods. Borrowing terminology from the physics community, we refer to such an idealized situation throughout this work as ``non-interacting'' because, for such a situation, there are no influences that act in a way that distinguishes among members of different subgroups when it comes to their neighborhood placement. 
 
Although we would not expect such a ``non-interacting'' distribution in practice, this non-interacting model serves as an idealized null hypothesis when testing for segregation. The entropic indices, which we showed also used information theory to compare regional and neighborhood-level knowledge, then can fittingly be computed directly from this non-interacting model of segregation (Appendix \ref{multinomial and ent}). Specifically
\begin{equation}
\begin{aligned}
    D_i = -\frac{1}{t_i} \ln P\{t_i\},
\end{aligned}\label{newdiv}
\end{equation}
and
\begin{equation}
\begin{aligned}
    S = -\frac{1}{TE} \sum_i \ln P \{t_i\}.
\end{aligned}\label{newend}
\end{equation}
Thus, from a statistical point of view, these indices correspond to negative log-likelihoods of the non-interacting model of segregation and allow us to propose a straightforward interpretation for the entropic indices: \emph{they indicate how unlikely the observed neighborhood occupancies would be to occur through random groupings of individuals in a region}. 
When the $-\ln P\{t\}$ values are small, the probabilities are large, and hence the observed neighborhood occupancies are consistent with a lack of subgroup-dependent interactions. Conversely, when the $-\ln P\{t\}$ values are large, the observed neighborhood occupancies are highly unlikely to occur in a non-interacting situation. 
Finally, by summing over all neighborhoods in a region and normalizing by the maximum possible value, the entropy index provides a 0 to 1 measure on `how far from non-interacting' the neighborhood occupancies are collectively for a region.

A number of important conclusions follow from the above observations. First, a maximum entropy model constructed from even very little information can be quite meaningful. Second, the entropic indices correspond to the unlikelihood of the observed neighborhood occupancies under the null hypothesis of the non-interacting, multinomial, model of segregation. Traditional entropic indices are thus measures based on the presumption of the consistent \emph{absence} of subgroup-dependent interactions, regardless of regional composition, and in this sense can be considered compositionally invariant. We emphasize that this does not mean that the values of these indices are independent of the regional composition, but rather that the \emph{presumed subgroup-dependent interactions (or lack thereof) are compositionally invariant}. The absence of interactions which underlies the entropic indices is, of course, a highly artificial assumption which we remove in the proceeding section.

\section{Interacting Systems} \label{sec:interacting}
\subsection{Maximum Entropy Model Using the Compositional Distribution} \label{sec:Hmodel}
The assumption that residential segregation is absent of subgroup-dependent interactions --- what we have seen to be the underlying basis of common entropic indices --- is a substantial oversimplification. It is entirely possible that measurements of the entropy index for regions with vastly different forms of subgroup-dependent interactions could give similar index values by sheer coincidence. To overcome this lack of specificity, we shall follow the logic of the previous section but now use the entire observed compositional distribution, as opposed to merely the regional composition. We do this because, as we have demonstrated, the compositional distribution fully encapsulates segregation in demographic data which incorporates the aspatial and scale-invariant properties that all common segregation measures tend to satisfy. 

The resulting occupancy model for a neighborhood has some features in common with the non-interacting (multinomial) model of segregation and takes the form
\begin{equation}
\begin{aligned}
    P\{t\}= \frac{1}{Z_{t_i}} \frac{t_i!}{\prod_m t_{im}!} e^{-t_iH\{\pi\} },
\end{aligned} \label{dfft form}
\end{equation}
where $Z_{t_i}$ is an overall normalization constant that is dependent on the neighborhood size $t_i$, and $H\{\pi\}$ is a region-specific `segregation function' that is the same for all neighborhoods (Appendix \ref{App:Int1}). Note that, because this distribution is dependent on the neighborhood size, the observed compositional distribution for a given region must ultimately be generated through a neighborhood-size weighted average of this model of segregation. Thus, when the neighborhood sizes are specified, the function $H\{\pi\}$ has a one-to-one relationship with the compositional distribution for a region.

\subsection{Compositionally Invariant `Simplified Data': \emph{Compositional Behavior}} \label{sec:CD-xforms}
In a broad sense, $H\{\pi\}$ is a segregation function which captures the tendency for neighborhood compositions to be `avoided', being large when the probability of a composition is low and \emph{vice versa}, and has been called the `headache' function in related work on crowd physics (\cite{mendez2018density}). The function $H\{\pi\}$, however, is dependent on the regional composition and it is natural to ask whether this dependence can be removed. 

To this end, suppose that we have parameterized the headache function for one region, $H_1\{\pi\}$, and want to determine a headache function for a different region, $H_2\{\pi\}$. In the absence of any information regarding the second region, $H_1\{\pi\}$ is perhaps the most reasonable estimate for $H_2\{\pi\}$. Alternately, if the entire compositional distribution of the new region is known, then $H_2\{\pi\}$ can simply be computed directly. Such intuition is placed on a rigorous information theory basis using the \emph{principle of minimum cross-entropy} (MCE) (Appendix \ref{CE}). Through the use of MCE, the initial model, described by $H_1\{\pi\}$, can be adjusted to conform with a new regional composition while reducing entropy (uncertainty) only as much as absolutely necessary. Conceptually, then, MCE determines an `updated model' that minimizes the introduction of any additional assumptions. We note that MCE also forms the basis of the \textit{iterative proportional fitting} (IPF) approach employed by Elbers (2021\nocite{elbers2021method}) to determine the marginal components of segregation. However, unlike IPF, MCE can be performed analytically to determine formulas for how $H\{\pi\}$ is expected to change due solely to changes in the regional composition. This MCE procedure then allows us to determine the compositionally invariant aspect of a compositional distribution as follows.

From minimum cross-entropy, the change in the headache function due to knowledge of a new regional composition results in a correction that is linear in the $\pi_{im}$ (Appendix \ref{CVMCE}), so that $H\{\pi\}$ can be written as
\begin{equation}
\begin{aligned}
    H \{\pi\} = \sum_m \pi_{im} v_m + f\{\pi\},
\end{aligned}\label{H partioning}
\end{equation}
where $f\{\pi\}$ is a compositionally invariant segregation function and the $v_m$ are determined by the regional composition. 
Finally, the corresponding occupancy model is
\begin{equation}
\begin{aligned}
    P\{t\}= \frac{1}{Z_{t_i}} \frac{t_i!}{\prod_m t_{im}!} e^{-t_i \big(\sum_m \pi_{im} v_m +f\{\pi\} \big)},
\end{aligned} \label{dfft form2}
\end{equation}
or, in the binary case,
\begin{equation}
\begin{aligned}
    P(t_{im})  = \frac{1}{Z_{t_i}} \frac{t_i!}{t_{im}!(t_i-t_{im})!} e^{-t_i \big(\pi_{im} v_m +f(\pi_{im}) \big)},
\end{aligned} \label{dfft form binary}
\end{equation}
where $t_{im}$ is the population of the group of interest in a neighborhood of size $t_i$. Note that these models are exactly the same as those obtained using density-functional fluctuation theory (DFFT) through physics-based arguments which refer to $v$ as the `vexation' and $f$ as the `frustration' (\cite{mendez2018density}). The function $f\{\pi\}$ itself will then be referred to as the `frustration' and the phenomena which it describes, the compositionally invariant aspect of the compositional distribution, as the `compositional behavior'. We emphasize that although the partitioning of $H\{\pi\}$ ended up being rather simple, the relationship between the $f\{\pi\}$, the $v_m$, and the neighborhood sizes are non-linear in their effect on $P\{t\}$: this is not a simple partitioning of the compositional distribution into a sum of marginal and structural components.

To better interpret the meaning of the frustration $f\{\pi\}$ we note that, when there are no subgroup-dependent interactions, we can expect $f\{\pi\} = 0$ because the distribution of equation~\ref{dfft form binary} is then exactly mathematically equivalent to the multinomial form of equation~\ref{eq:multinomial}. Thus,  $f\{\pi\}$ indicates the extent of departure from non-interaction. Generally, frustrations with positive curvature tend to have lower values toward the center of their domain, indicating a tendency for integration. Conversely, frustrations with negative curvature favor extreme values of the composition, indicating tendency for segregation. Of course, being determined directly from data, $f\{\pi\}$ can take forms of arbitrary complexity corresponding to the broad range of compositional behaviors that can arise from compositional distributions.

We further note that $f\{\pi\}$ does not, at least without further analysis, elucidate \emph{why} such compositional behaviors occur and instead it should be considered the best representation of subgroup-dependent interactions that is possible with demographic data satisfying the aspatial, scale invariant, and compositionally invariant properties. These interactions captured by $f\{\pi\}$ then represent a conglomeration of individual preferences/aversions, governmental policy, geographical features, \emph{etc}. To summarize, given a compositional distribution,  $f\{\pi\}$ captures the `behavior' that cannot be explained by the regional composition. 

Crucially, with this framework in place, it is now possible to compare segregation between regions which differ in their regional compositions without the need to resort to `non-interaction' as a baseline for comparison. Specifically, given the compositional distribution for a given region, Region~1, one can adjust the values of the vexation variables $v_m$ in equation \ref{dfft form2} or \ref{dfft form binary} to transform the compositional distribution to what would be expected for the regional composition of a second region, Region~2. The transformed compositional distribution of Region~1 can then be compared directly with the observed compositional distribution of Region~2, allowing comparison in the absence of differences in regional compositions. Moreover, to allow direct comparison of the segregation of multiple regions at once, equation~\ref{dfft form2}  or \ref{dfft form binary} can be used to transform the compositional distributions of each region to the same regional composition. Finally, to remove the effects of neighborhood size as well, we can similarly determine the compositional distribution that would arise if all regions had the same neighborhood sizes. In this work, we will always compare (binary) compositional distributions after transformation to a regional composition of $\pi_m$=50\% White with all neighborhoods having a size of $t_i$=1,000. We emphasize that this choice is specific to our situation of binary segregation for block groups, which are often around 1,000 in size, and that for other applications the most reasonable values to choose could differ.

By determining such `standardized compositional distributions' (SCDs) that result from the above process, the $v_m$ and $t_i$ in equation \ref{dfft form binary} become fixed and thus any remaining degrees of freedom are determined by the data-driven $f\{\pi\}$. This is useful in a number of ways. First, plotting the standardized compositional distribution constitutes a direct illustration of structural segregation because marginal differences have been removed. As a corollary, the standardized compositional distributions and anything calculated from them will be compositionally invariant. Specifically, as SCDs are compositional distributions, we retain the ability to compute `standardized' values for traditional indices and hence retain the conceptual simplicity of comparing index values. We refer to these compositionally invariant versions of traditional indices as \emph{SCD segregation indices}. 

\section{Standardized Compositional Distributions: NYC and Chicago \emph{vs.} LA} \label{sec:SCDs}

\begin{figure}[hbt!]
    \includegraphics[width=6.6in,keepaspectratio]{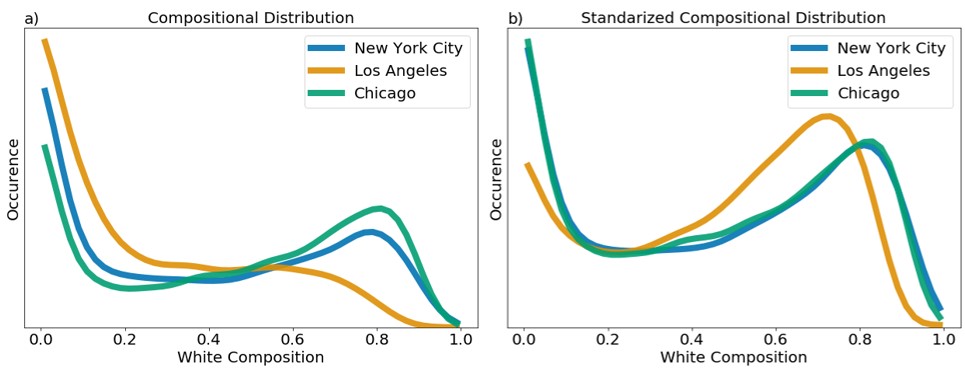}
    \caption{\textbf{White/non-White compositional distributions of the three largest metropolitan areas in the United States}: (a) observed compositional distributions, (b) standardized compositional distributions. The observed compositional distributions appear quite distinct for New York City, Los Angeles and Chicago. However, after standardizing the regional compositions 
 to 50\% White and neighborhoods to size 1,000, New York City and Chicago exhibit nearly identical White/non-White segregation patterns. (See Appendix~\ref{num_op}-\ref{non-smoothed} for numerical details.)}
    \label{fig:LargestMetros}
\end{figure}

To illustrate the power of this approach, consider the White/non-White compositional distributions of the three largest metropolitan areas in the U.S. (New York City, Los Angeles and Chicago), as shown in Figure \ref{fig:LargestMetros}. Looking at just the initial, unadjusted, compositional distributions (Figure \ref{fig:LargestMetros}a), the three cities appear distinct, with only some hints that New York City and Chicago most resemble each other. However, once we account for marginal differences and consider the standardized compositional distributions (Figure \ref{fig:LargestMetros}b), it becomes abundantly clear that White/non-White structural segregation in Los Angeles is quite distinct from New York City and Chicago and, further, that those of New York City and Chicago are nearly identical. We also note that, despite the clearly distinct nature of the Los Angeles compositional behavior, all three cities show the same general trends in their SCDs. This is especially true when they are compared to what we would expect in the absence of subgroup-dependent interactions: a sharply peaked binomial distribution at 50\% White. Compared to such a non-interacting binomial distribution, all three metropolitan areas show relatively broad distributions with a peak near 0\% White composition, a broad minimum near 20\% White composition, a much more well-defined peak in the 60-80\% White range, and a steep drop at compositions above 90\% White.

With access to both the original compositional distributions and, now, the standardized compositional distributions, we consider the additional insights to be gained by using these distributions to compute `standardized' segregation index measures (Table 2). Before considering the main results, we emphasize that marginal variables can be accounted for sequentially, such as, for example, removing differences resulting from regional compositions while retaining differences due to neighborhood sizes. As described above in section~\ref{sec:interacting}, the observed compositional distribution can be computed by forming a size-weighted average of model distributions using the known neighborhood sizes in a region. A compositional distribution can then be predicted for a different regional composition by determining the vexation $v_m$ values that would lead to the predicted compositional distribution having a desired regional composition.  Comparison of the second column of the table, which keeps neighborhood size differences but sets the regional composition to 50\% White, with the third column, which sets the regional compositions to 50\% White \emph{and} all neighborhood sizes to 1,000, shows that the impact of neighborhood sizes is quite small and suggests that the regional composition tends to be the dominant marginal effect. 

From their original compositional distributions (Table 2, first column), New York City, Los Angeles and Chicago have distinct White compositions (43.3\%, 28.5\%, 50.2\%, respectively) and all demonstrate similar White/non-White dissimilarity index values (0.561, 0.523, 0.515). It is important to recall that the binary dissimilarity index has traditionally been considered to be compositionally invariant, so these values would be expected to remain constant when we adjust the regional compositions to 50\% White. However, when we compute the dissimilarity index from the standardized compositional distributions (Table 2, third column), which have a 50\% White regional composition, we find dissimilarity index values (0.534, 0.432, 0.521) that change noticeably, with a particularly notable change for Los Angeles. This larger change for Los Angeles is due to its regional composition undergoing the largest adjustment between the original compositional distribution and its standardized counterpart. Moreover, despite Los Angeles having a traditional dissimilarity index value that is similar to New York City and Chicago, indicating similar segregation in the evenness dimension, we find this to be an artifact of differences in regional compositions and not representative of similar structural segregation. The results obtained from the SCD dissimilarity index values, those computed from standardized compositional distributions, are reasonable: prior studies have noted that heterogeneity in regional segregation reflects variations in segregation histories, the typical age of housing, and the relative size of Latino and Asian populations (\cite{iceland2013sun}; \cite{charles2003dynamics}), variables which distinguish Los Angeles from the Northern cities. This result, however, would have been obscured in traditional analysis of White/non-White segregation for these metropolitan areas. 

Finally, we consider the impact of our analysis on the isolation index for the non-White subgroup. The resulting index values then represent the isolation for the regions \emph{if} they happened to have the same overall compositions and neighborhood sizes. We first note that, in the absence of subgroup-dependent interactions (non-interacting case), a region with a standardized composition of 50\% White should have an isolation index value of 0.5, because a non-White individual would be equally likely to encounter a non-White and White individual. An isolation index value of greater than 0.5, then, would indicate more intragroup encounters than expected for a non-interacting system and, consequently, the presence of structural segregation. Thus, through the adjustment afforded by SCDs, even measures of exposure, which traditionally have a clear dependence on the regional composition, can be used as proxy measures of structural segregation. The third column of Table 2 indeed shows that the SCD isolation index suggests that New York City and Chicago exhibit very similar structural segregation while Los Angeles is comparatively less segregated, entirely consistent with the SCD dissimilarity index results. As a final note, although the isolation index is generally not symmetric between groups, it \textit{is} symmetric when the groups are of the same size in the region. Hence, as the SCD is chosen to have a regional composition that is 50\% White, the SCD isolation~index is the same whether it is measured from the perspective of the White or non-White group.

\section{Historical Trends: Isolating National Structural Segregation} \label{sec:Historical}
We found the SCDs of the three largest metropolitan areas in the U.S. to have similar overall patterns in 2020: a peak around 0\% White, a peak in the 60-80\% White range, and a steep drop above $\sim$90\% White. As noted above, this demonstrates a clear and consistent departure from expectations given an absence of subgroup-dependent interactions, which would yield an SCD characterized by a sharply peaked binomial distribution centered at 50\% White. These results then suggest that determination of a representative \emph{national standardized compositional distribution} may give useful insights into broad trends in structural segregation over time.

\begin{figure}[hbt!]
    \includegraphics[width=6.6in,keepaspectratio]{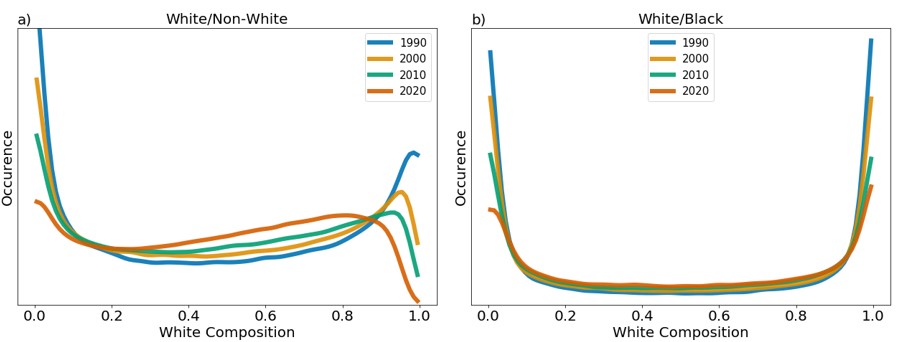}
    \caption{\textbf{Historical evolution of national standardized compositional distributions}: (a) White/non-White, (b) White/Black.  The national SCDs are calculated using all metropolitan statistical areas in the United States for the corresponding decade. The pattern of the White/non-White standardized neighborhood compositional distribution has changed substantially over the last 30 years, with a near disappearance of the tendency for all-White neighborhoods, and integrated neighborhoods becoming quite favored. The White/Black standardized distribution has retained a tendency for neighborhoods at the extremes, but the strength of this tendency has consistently and significantly decreased. (See Appendix~\ref{num_op}-\ref{non-smoothed} for numerical details.)}
    \label{fig:White-non-White}
\end{figure}

Extraction of a representative national SCD can be performed using a statistically rigorous maximum-likelihood estimation (MLE) procedure that accounts for variations in regional sizes and results in a compositional behavior, and hence SCD, most representative of the entire nation (Appendix \ref{multiregion}). Applying this procedure to decennial U.S. census data from 1990 through 2020 yields the historical White/non-White and White/Black national SCD trends shown in Figure~\ref{fig:White-non-White}. This reveals that the structural segregation tendencies observed in the previous section, for White/non-White disassociation in the largest metropolitan areas in 2020, are a rather recent development. For example, the 1990 White/non-White SCD indicates a strong tendency for both fully non-White and fully White neighborhoods, which has changed precipitously. By 2020, the tendency for fully non-White neighborhoods has become comparable to integrated neighborhoods in the 60-80\% White range, and fully White neighborhoods have become extremely disfavored. For White/Black segregation (Figure~\ref{fig:White-non-White}b), the general shape of the national SCDs has remained highly symmetric around 50\% White over time, with highly segregated neighborhoods remaining most favored. However, there has been a persistent and significant drop in the tendency for such highly segregated neighborhoods and a corresponding increase in the prevalence of integrated neighborhoods, demonstrating a historical reduction in structural White/Black segregation as well.

\begin{figure}[hbt!]
    \includegraphics[width=6.5in,keepaspectratio]{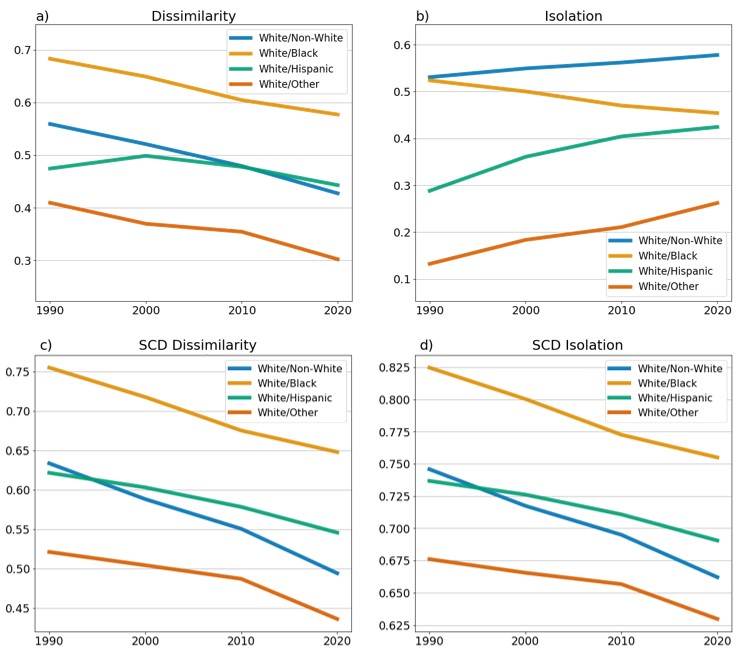}
    \caption{\textbf{Historical trends in national segregation indices:} (a-b) traditional indices, (c-d) SCD indices. For each decade and binary segregation comparison, the traditional index values (a-b) represents a size-weighted average of regional indices for (a) the dissimilarity index and (b) the isolation index. For the SCD counterparts (c-d), a national standardized compositional distribution (neighborhoods of size 1,000 and a regional composition that is 50\% White) was obtained and the respective index values were calculated. Using the national SCDs, we find that structural segregation has decreased over time in all cases, with remarkably similar trends with both the SCD dissimilarity and SCD isolation indices.}
    \label{fig:White-other indices}
\end{figure}

As in the previous section, it is also useful to summarize the national SCDs by calculating the corresponding standardized segregation indices. Figure \ref{fig:White-other indices} (upper panels) shows the historical nationally-averaged dissimilarity and isolation trends for a wide range of binary comparisons (White/non-White, White/Black, White/Hispanic, White/Other), as would be calculated traditionally, and Figure \ref{fig:White-other indices} (lower panels) shows the SCD counterparts, which account for historical shifts in demographics through our standardization of the compositional distributions. It is noteworthy that only the White/Black comparison demonstrates consistent historical trends with traditional and SCD indices, specifically, a persistent reduction in segregation. This results from the Black subgroup being the only minority group which has remained a roughly constant proportion of the population throughout the decades, meaning that historical comparisons with traditional indices are roughly compositionally invariant. With the SCD indices removing differences in historical demographics, not only do we retain this downward trend for White/Black segregation but we also discover nearly-linear downward trends for all considered binary comparisons. This demonstrates that not only has White/Black segregation decreased, but that there has been a persistent decrease for White/non-White, White/Hispanic, and White/Other structural segregation as well. 

Perhaps the most striking result, however, is that the historical trends for the SCD dissimilarity and isolation indices are always virtually identical. These results are remarkably distinct from those obtained with traditionally computed indices, which tend to indicate vastly different trends depending on which index is used. For example, prior studies found no substantive change in White/Hispanic segregation using the dissimilarity index whereas the isolation index reflected the substantial increase in the Hispanic population (\cite{logan2022metropolitan}). We find, instead, that after accounting for the increase in the Hispanic population, White/Hispanic structural segregation presents no disageement between the dissimilarity and isolation indices. Hence, following the proper removal of marginal differences, Gorard's mantra that ``almost any index would do'' (2007:672\nocite{gorard2007does}) may now well apply to indices of different segregation dimensions. 

\section{Regional Comparisons}
\subsection{Predictions: Index Dependence on the Regional Composition} \label{sec:IXA}

\begin{figure}[hbt!]
    \includegraphics[width=6.5in,keepaspectratio]{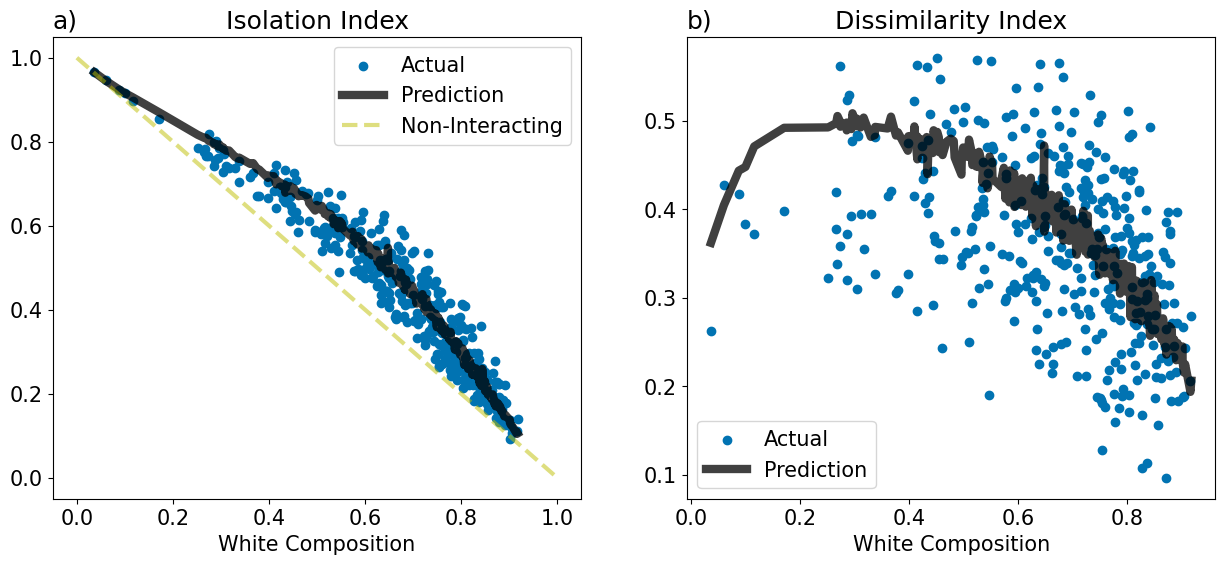}
    \caption{\textbf{Dependence of (a) White/non-White isolation index and (b) dissimilarity index on the regional White composition for the United States:} observed values (blue dots), predictions from national compositional behavior (grey curve), isolation-index expectations from non-interacting model (yellow dashed line). (The dissimilarity index would consistently be near 0 for the non-interacting case.) The observed values for both indices differ significantly from the non-interacting expectation, indicating the presence of structural segregation. In both cases, the index values predicted from the national SCD capture the deviations from the non-interacting case well, indicating that the national norm gives a reasonable baseline description of segregation in the U.S.
    (The slight roughness of the predicted curve \ arises from variations in neighborhood sizes among the regions.) }
    \label{fig:White-compositional dependence}
\end{figure}

Not only are national standardized compositional distributions useful for demonstrating how national trends develop over time, the underlying compositional behaviors can also be used to determine the dependence of indices on the regional composition. Figure \ref{fig:White-compositional dependence} shows White/non-White isolation and dissimilarity indices of all U.S. metropolitan statistical areas in 2020 along with predictions using the 2020 White/non-White national compositional behavior. Specifically, the predictions are formed by creating an expected compositional distribution for each region by combining its specific marginal variables, regional composition and neighborhood sizes, with the national $f\{\pi\}$ and then computing the index values.

In the non-interacting case, the isolation index is expected to have a direct linear dependence on the regional White composition. This appears to be roughly true of the observed data, but with a systematic deviation in the 20\% to 80\% White range. This deviation from the expected linear behavior is captured quite well using the national compositional behavior and the regional marginal variables (Figure \ref{fig:White-compositional dependence}a), suggesting that structural segregation is more consistent across the U.S. than comparison to a non-interacting model of segregation would imply. 

Perhaps more interesting are the results for the dissimilarity index (Figure \ref{fig:White-compositional dependence}b), which is a measure widely considered to be both compositionally and scale invariant (\cite{frankel2011measuring}). We would then expect that as we parameterize our model with different marginal variables, the dissimilarity index predictions should remain constant. Despite this, our model finds a clear dependence of the dissimilarity index on the regional composition: high dissimilarity index values for regions around 30\% White and low dissimilarity index values for predominantly White regions. These changes in the dissimilarity index span as much as 0.3 due solely to changes in the regional composition. Crucially, this trend is borne out by the dissimilarity index values that would be traditionally computed for the regions, with the observed and nationally predicted dissimilarity index values having a Pearson correlation coefficient of over 0.5. Not only does this demonstrate that the dissimilarity index is a poor measure of structural segregation if differences in regional compositions are ignored,  but it also demonstrates that the dependence of the dissimilarity index on the regional composition is largely predictable.

\subsection{Counterfactuals: Geographical Distributions of High \& Low Structural Segregation} \label{sec:IXB}

To determine differences in structural segregation between regions, the most familiar approach would be incorporating compositional behavior to create index counterfactuals. The idea is to generate an expectation for an index, and then to consider the discrepancy with observations to be the result of differences in structural segregation. This is the essence of the approach employed by Elbers~(2021\nocite{elbers2021method}). Here, we will use expectations from the previous section, which are predictions for regional indices based on region-specific marginal variables and a national compositional behavior, to explore which regions are more or less segregated with respect to the national norm. 

\begin{figure}[hbt!]
    \includegraphics[width=6.5in,keepaspectratio]{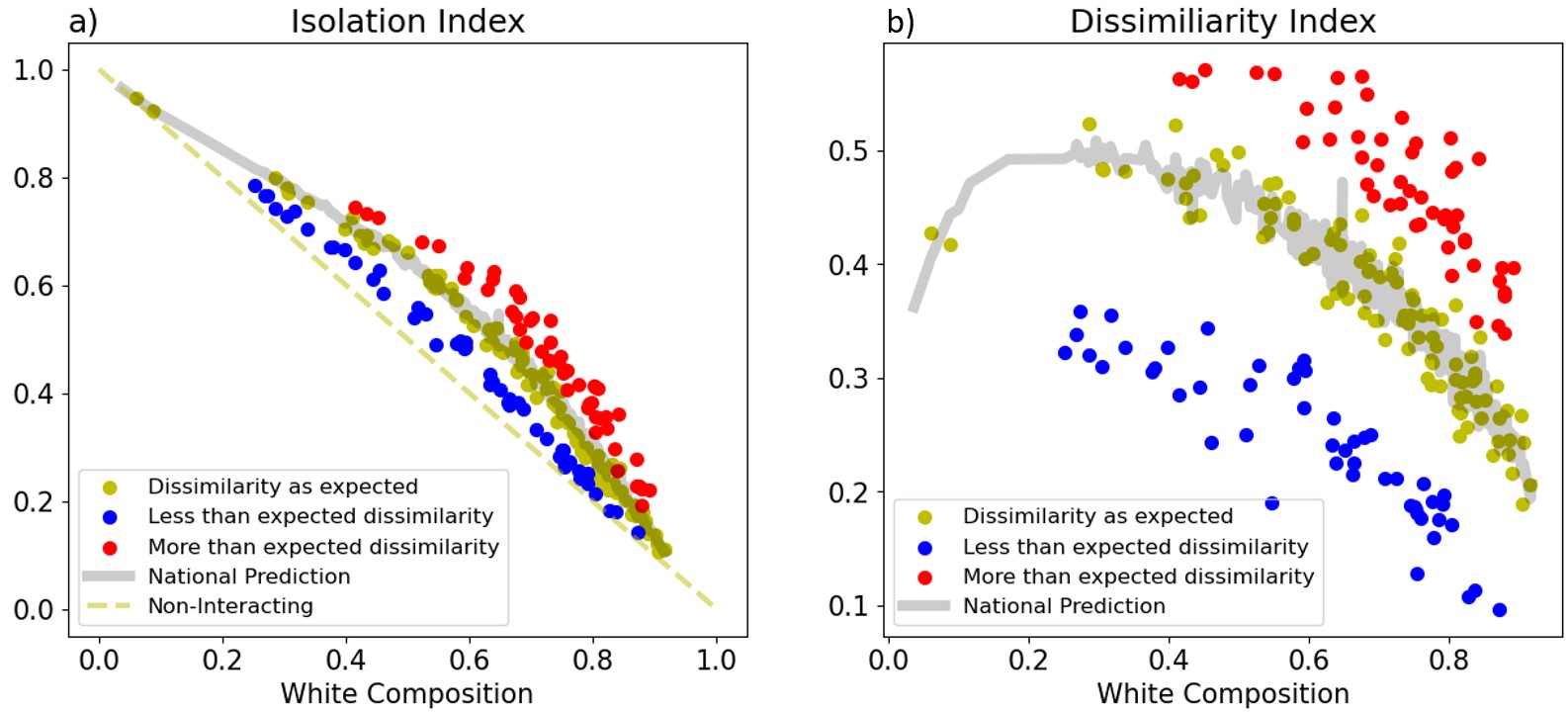}
    \caption{\textbf{Consistency in capturing structural segregation between indices:} same as  Figure~\ref{fig:White-compositional dependence} but displaying only the 100 regions with the best prediction for the dissimilarity index (yellow dots) and the 50 regions with the most over-estimated (red dots) and under-estimated (blue dots) dissimilarity indices. The isolation index values for each of these three groups of regions follow the same trends relative to the national norm expectation for the isolation index, indicating that the differences from the norm expectations are reflective of regions of stronger or weaker structural segregation.}
    \label{fig:White-compositional dependence2}
\end{figure}

We begin with the dissimilarity index, which is often employed to compare segregation between different regions. Figure \ref{fig:White-compositional dependence2}b highlights the 100 regions whose White/non-White dissimilarity index is best predicted using the national compositional behavior (yellow circles), as well as the 50 regions with the greatest over-estimation (red circles) and under-estimation (blue circles) of the dissimilarity index. The latter two groups then constitute regions which are, respectively, most and least structurally segregated relative to the national norm. To further support this identification, Figure \ref{fig:White-compositional dependence2}a displays the isolation index for the above three sets of regions. We find that the 100 best predicted regions by their dissimilarity index also had their isolation index predicted very well: nearly 80\% have their isolation indices predicted to within 0.02, and all of these regions, representing over one-fourth of all metropolitan statistical areas in U.S., have isolation indices predicted to within 0.043, indicating near perfect agreement. This agreement between the nationally predicted dissimilarity and isolation index also tends to hold well for other binary comparisons such White/Black and White/Hispanic (Appendix \ref{NatPredReg}).

Moreover, we note that our dissimilarity results underscore the fact that ignoring the impact of the regional composition on the dissimilarity index can lead to misleading results. Specifically, some of the regions we have identified as having amongst the highest structural segregation relative to the national norm (red circles occurring above 80\% White composition) actually have \emph{lower traditional dissimilarity index values} than some of the regions we have identified as being amongst the least structurally segregated (blue circles near 20\% White composition). 

\begin{figure}[hbt!]
    \includegraphics[width=7in,keepaspectratio]{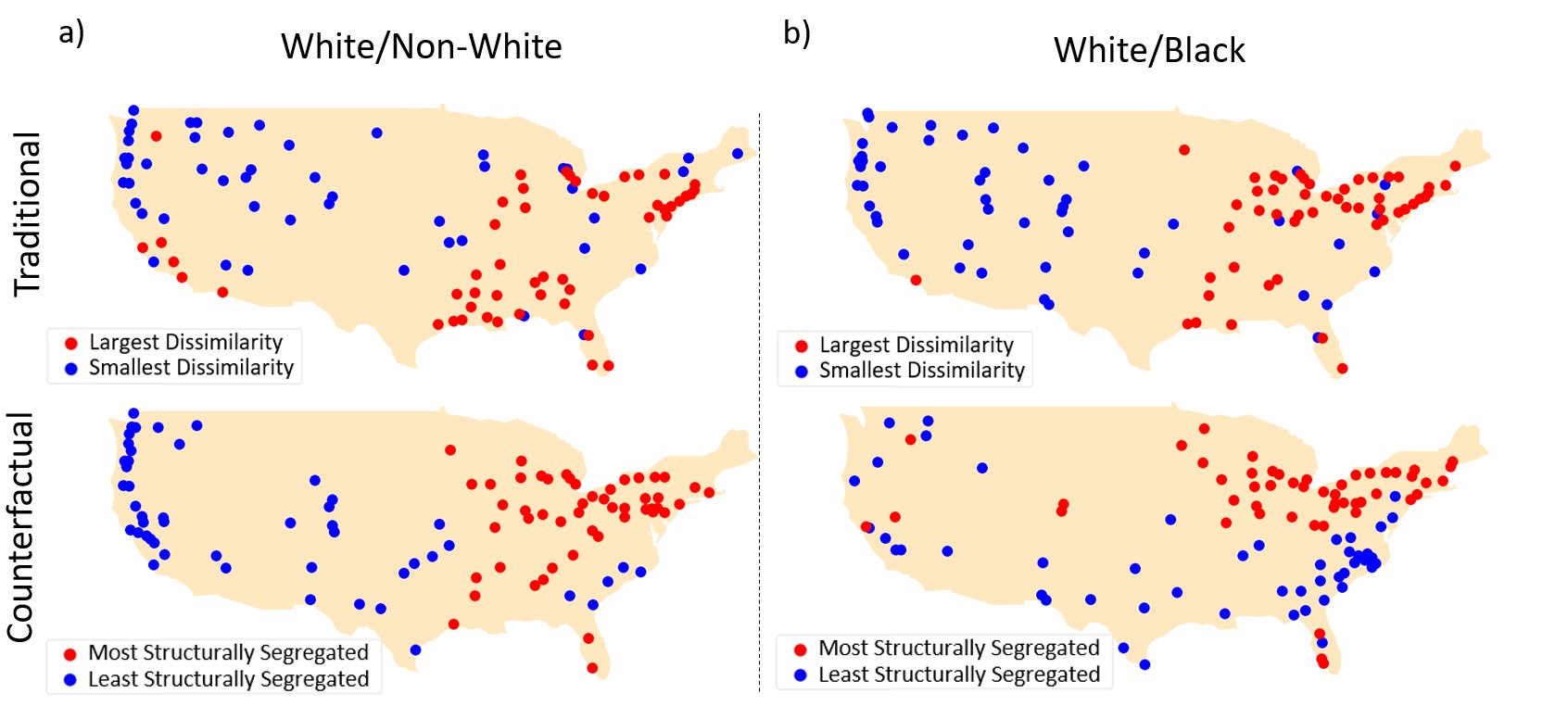}
    \caption{\textbf{Spatial clustering of extremes of segregation:} least (blue) and most (red) segregated metropolitan regions as determined using the traditional dissimilarity index (upper panels) and the proposed comparison relative to the national norm (lower panels) for White/non-White (a) and White/Black (b) segregation. The counterfactual accounts for differences in overall regional compositions and neighborhood sizes, and greatly clarifies broad geographical trends. Specifically, strong White/non-White segregation consolidates to the Northeast and Midwest and the least segregated regions consolidate to the Southeast and Western U.S. Similarly, White/Black segregation consolidates to the Northeast, with the least structurally segregated regions shifting dramatically to the Southeast.}
    \label{fig:White-Non-White Dissimilarity USA}
\end{figure}

 Next, having identified those metropolitan areas which are most and least structurally segregated relative to the national norm, we consider their geographical locations. Figure~\ref{fig:White-Non-White Dissimilarity USA}a shows the geographical distribution of the fifty least and most segregated metropolitan areas in the U.S., as determined either using traditional values for the White/non-White dissimilarity index (Figure~\ref{fig:White-Non-White Dissimilarity USA}a, upper panel) or using the the regions identified above by counterfactuals underpinned by a national compositional behavior (Figure~\ref{fig:White-Non-White Dissimilarity USA}a, lower panel). We find that traditional analysis suggests that the most segregated regions are predominantly found in the eastern half of the United States, and that the least segregated regions are quite well dispersed (upper panel). After accounting for marginal differences, however, we find a greater consolidation of the most structurally segregated regions to the Midwest and Northeastern U.S., and a consolidation of the least structurally segregated regions to the Western U.S. and the Southeast. Thus, White/non-White segregation exhibits a significantly stronger geographical coherence when marginal differences are removed. 
 
A similar picture emerges for the more commonly considered White/Black segregation comparison, for which a plot along the lines of Figure~\ref{fig:White-compositional dependence2} can be made (Appendix \ref{NatPredReg}) and the geographic distributions determined (Figure~\ref{fig:White-Non-White Dissimilarity USA}b).  We find that the traditional White/Black dissimilarity index (upper panel) also suggests an East/West division between the least and most segregated regions, though it must be noted that the identified regions in the West tend to have very few Black individuals. After using White/Black national compositional behavior and accounting for marginal differences, a substantial change in the geographic distributions is found. Both the least and most structurally segregated regions are found in the Eastern half of the United States, with the most structurally segregated regions tending to occur in the Northeast and the least structurally segregated regions in the Southeast. These are regions that tend to have substantial Black populations but have experienced vastly different histories, indicating that White/Black segregation has remained structurally distinct along historical geographic lines. 
 
 \subsection{Indexless Analyses: Segregation Dependent on Minority Diversity} \label{sec:indexless}
 
So far we have used national compositional behaviors and relied on traditional segregation indices for analysis, but compositional behaviors can be compared directly without introducing the additional assumptions inherent in traditional indices. This section first shows how to quantify the accuracy with which a compositional behavior represents a compositional distribution and, then, demonstrates how this quantification can be used to cluster regions into groups consisting of similar compositional behavior.

To determine how well a particular compositional behavior represents a compositional distribution, the natural choice is to use cross-entropy. Specifically, this measure can be written in a form that we refer to as the `generalized M index',
\begin{equation}
\begin{aligned}
    M_g = -\frac{1}{T} \sum_i \ln P (\{t_i\} | f\{\pi\}, v_m),
\end{aligned}
\end{equation}
where $P (\{t_i\} | f\{\pi\}, v_m)$ is the expected model distribution based on equation \ref{dfft form2} given a frustration function $f\{\pi\}$ (using $f\{\pi\} = 0$ recovers the traditional $M$ index) and a set of vexation values $v_m$ (to obtain the correct regional composition). Finally, $T$ is the total population of the region. 

Whereas the traditional $M$ index determines how poorly a non-interacting model represents observations (section~\ref{sec:logliklihood}), the $M_g$ index does the same for any choice of model distribution.  In this sense $M_g$ is a `relative' index, it requires a choice of model distribution and so typically does not provide an absolute scale to determine regions of greatest or least segregation. Instead, $M_g$ can be used to determine those regions which are well-represented by the same segregation model and, hence, can be used to cluster regions in the United States by the similarity of their compositional behavior. To accomplish this, we follow a standard clustering algorithm that starts by randomly assigning  regions into a fixed number of `clusters' and then moving each region to the cluster whose representative compositional behavior best reflects the observations in that region. More precisely, for each cluster, we determine its representative compositional behavior and then move each region to the cluster which yields the smallest $M_g$ value. We then perform this calculation iteratively until no region needs to be reassigned to a new cluster. The final result is a set of clusters for which every region belongs to the cluster with the closest representative compositional behavior.

\begin{figure}[hbt!]
    \includegraphics[width=7.1in,keepaspectratio]{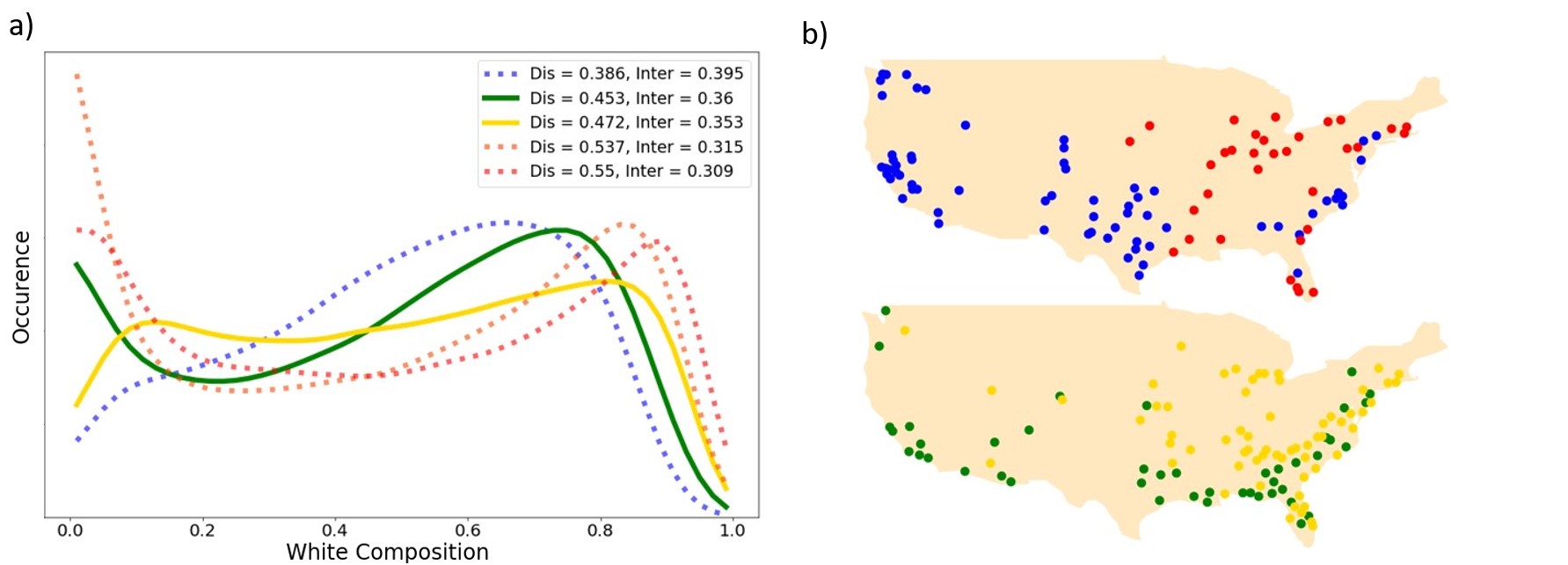}
    \caption{\textbf{Indexless clustering of metropolitan regions of 25-75\% overall White composition based on White/non-White segregation:} (a) standardized compositional distributions for each cluster and (b) geographical distribution of the least/most segregated clusters (blue/red, upper panel) and two clusters of intermediate segregation with very similar SCD index values (green/yellow). The least and most segregated clusters are largely consistent with the geographical results obtained using the national norm counterfactual approach (Figure~\ref{fig:White-Non-White Dissimilarity USA}a). The formation of two distinct, intermediately segregated clusters (green and yellow) with very similar SCD index values yet distinct geographic distributions suggests the existence of nuanced distinctions between these that may not captured with traditional indices.}
    \label{fig:clustered SCD}
\end{figure}

Figure~\ref{fig:clustered SCD} shows the result of applying this algorithm for White/non-White data using 5 clusters. For this analysis, we consider all metropolitan areas that have a substantial, but not overwhelming, regional White composition (between 25\% and 75\% White). The representative SCDs and geographical distributions of the resulting clusters are shown in Figures~\ref{fig:clustered SCD}a~and~\ref{fig:clustered SCD}b, respectively. We emphasize that these results stem from a highly automated process that represents an unbiased grouping of metropolitan areas into clusters by their compositional behavior. This procedure did not use any geographical information about the locations of the regions and did not rely on any traditional index values. Nevertheless, we find clusters with substantial geographical coherence, which demonstrates, unsurprisingly, that regions with close spatial proximity tend to have similar structural segregation. For example, the grouping into five clusters identifies two clusters at the extremes of segregation (red/blue, Figure~\ref{fig:clustered SCD}) that have spatial structures highly aligned with our counterfactual approach of the previous section. 

The clusters  for `intermediate' segregation, however, uncover substantially more nuanced trends. Specifically, two clusters (green and yellow) arise which have very similar SCD dissimilarity values (0.453 and 0.472, respectively) and SCD interaction index values (0.360 and 0.353) but clear differences in their SCD. For example, the SCD for the green cluster indicates more than twice the tendency for entirely non-White neighborhoods than does the yellow cluster but, also, less tendency for White dominated neighborhoods. These opposing trends lead to deceptively similar index values and suggest the existence of nuances of segregation that may be missed with traditional analyses, and so we now investigate these trends in more detail.

In an attempt to determine the demographic origin of the distinct compositional behaviors in the green and yellow clusters, we begin with the average values of traditional indices for these clusters. The green and yellow clusters show very similar dissimilarity index values on average ($0.397\pm 0.012$ and $0.392\pm 0.005$, respectively) and only a mild difference in the isolation index values ($0.637\pm 0.012$ and $0.726\pm 0.006$) which can largely be explained by the modest difference in the average White composition ($0.527\pm 0.020$ and $0.652\pm 0.010$). (The quoted uncertainties represent the uncertainty in the mean of the respective value.) In addition, consideration of a more detailed breakdown of the non-White group also finds rather modest differences between the clusters, with the composition of the non-White population being made up of $34\pm 4$\% and $40\pm 2$\% Black, $42\pm 4$\% and $34\pm 2$\% Hispanic, and $23\pm 2$\% and $25\pm 1$\% Other, respectively. These results suggest some minor differences between the clusters, but do not identify a clear distinguishing factor.

\begin{figure}[hbt!]
    \includegraphics[width=7in,keepaspectratio]{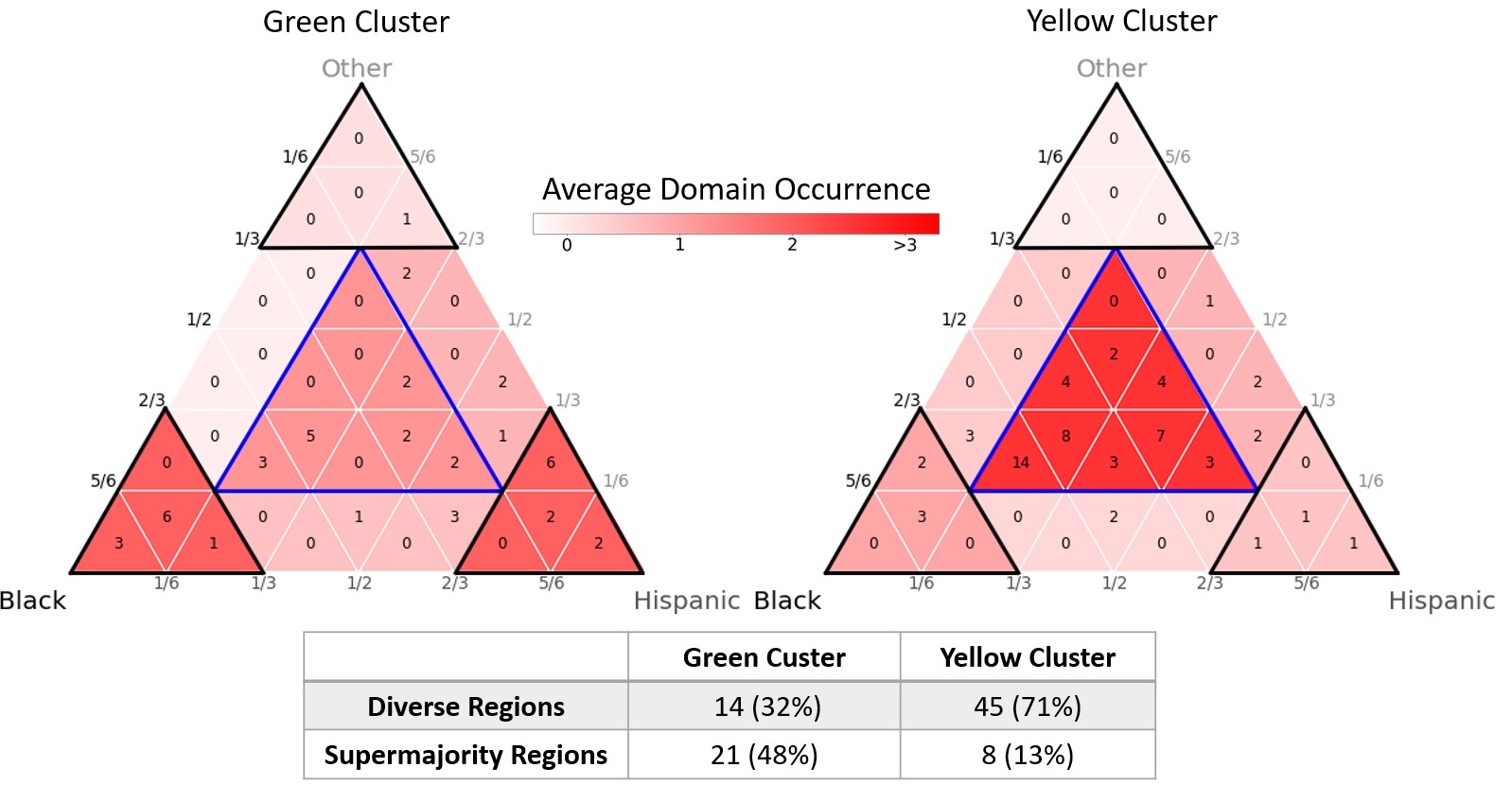}
    \caption{\textbf{Composition of the non-White population for clusters of intermediate segregation from Figure~\ref{fig:clustered SCD}:} Ternary plots for the green and yellow clusters counting supermajority regions, in which one minority group constitutes more than 2/3 of the non-White population (black triangles), and diverse regions, where all minority groups constitute between 1/6 and 2/3 of the non-White population (blue triangle). The numbering of individual triangles indicates how many regions fall within the non-White composition outlined by the triangle, and shading indicates the average triangle value over a domain (blue triangle, one of three black triangles, one of three trapezoids). The data suggest that a distinguishing factor between the green and yellow clusters is the \emph{diversity} of the non-White group.}
    \label{ternary}
\end{figure}

What \emph{does} distinguish the clusters, however, is the diversity of the non-White group, shown in Figure \ref{ternary}. We find that over 70\% of the regions in the yellow cluster have a non-White population that is `diverse' (consisting of no less than 1/6 and no more than 2/3 of each of the Black, Hispanic, and Other subgroups), whereas only 13\% of these regions have a non-White population with a `supermajority minority' (consisting of more than 2/3 of Black, Hispanic, or Other). In stark contrast, for the green cluster, only 32\% of the regions have a `diverse' minority population and nearly half, 48\%, of these regions have a `supermajority minority' population.

\begin{figure}[hbt!]
    \includegraphics[width=7in,keepaspectratio]{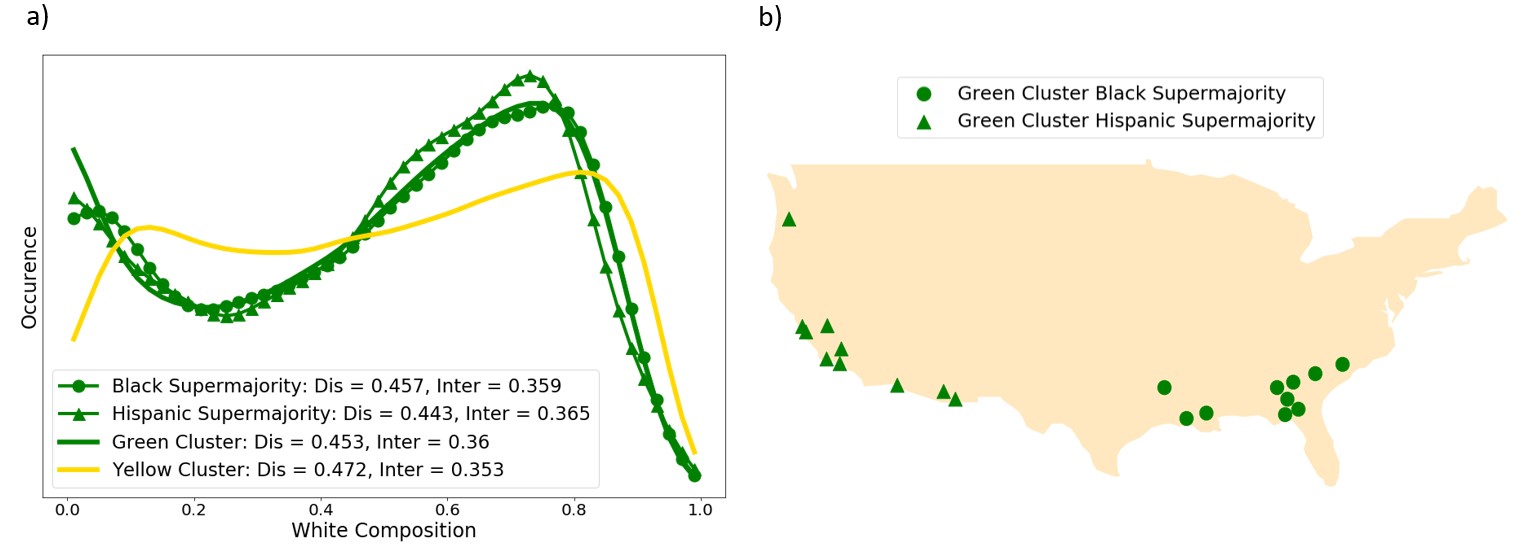}
    \caption{\textbf{Subclusters of supermajority minority regions by subgroup for the green cluster from Figure~\ref{fig:clustered SCD}}: (a) SCDs  and (b) geographic distribution for supermajority minority clusters dominated by Black (green circles) and Hispanic (green triangles) populations.  The strong agreement between the SCDs regardless of the dominant minority subgroup indicates that White/non-White structural segregation for the green cluster is indeed independent of the identity of the dominant minority. Furthermore, both of these subclusters occur in locations with a long-established historical presence of the respective minority group, suggesting that the long-term presence of a dominant minority subgroup leads to a particular form of structural White/non-White segregation. (See Appendix~\ref{num_op}-\ref{non-smoothed} for numerical details.)}
    \label{fig:SCD supermajority}
\end{figure}

These results regarding the diversity of the non-White group appear rather surprising, suggesting that a determining factor in White/non-White segregation may not be the dominance of a particular minority subroup but the overall diversity of the non-White population. To confirm that structural segregation is indeed independent of the identity of the supermajority minority subgroup, Figure \ref{fig:SCD supermajority}a compares the standardized compositional distributions for the subset of regions in the green cluster that have a supermajority Black or Hispanic minority population, respectively. The results indeed show nearly identical compositional behaviors regardless of whether the dominant minority subgroup is Black or Hispanic. It appears from these SCDs that the presence of any dominant minority leads to a tendency for entirely non-White neighborhoods (ethnic enclaves) and a tendency for fewer White dominated neighborhoods, possibly because the non-dominant minorities disperse throughout the region.

The geographic distribution of the Black and Hispanic supermajority regions also demonstrates a striking spatial structure (Figure \ref{fig:SCD supermajority}b), with the supermajority Black regions being entirely in the South-East and the supermajority Hispanic regions being along the West Coast. Future work should explore whether the similarities among these Black and Hispanic supermajority minority regions is a recent trend or if these similarities have held historically --- the latter would demonstrate that White/non-White structural segregation can evolve in similar ways for regions of vastly different geographical locations and demographics.

\section{Forecasting: how marginal differences mask reduction in structural segregation} \label{sec:forecast}

Another important aspect of our approach is that it can enable the incorporation of structural segregation into forecasting. Section~\ref{sec:CD-xforms} showed how compositional distributions change in response to variations in marginal variables. Then, section~\ref{sec:IXA} demonstrated that a national compositional behavior can lead to accurate predictions of index values for different regions. This suggests that, provided with future estimates of a regions's overall composition, distribution of neighborhood sizes, and a reasonable compositional behavior, we could forcast that region's compositional distribution and corresponding segregation indices. There are currently demographic models for projecting large-scale compositional changes (\cite{hauer2019population}). Moreover, as we demonstrated above, neighborhood sizes tend to constitute only a minor effect and so they only need to be projected roughly or, alternatively, can be assumed not to change. Finally, our results for the historical trends in the national standardized compositional distribution (Figure~\ref{fig:White-non-White}) suggest that some reasonable historical extrapolation of the national compositional behavior from historical data should be possible.

\begin{figure}[hbt!]
    \includegraphics[width=6.75in,keepaspectratio]{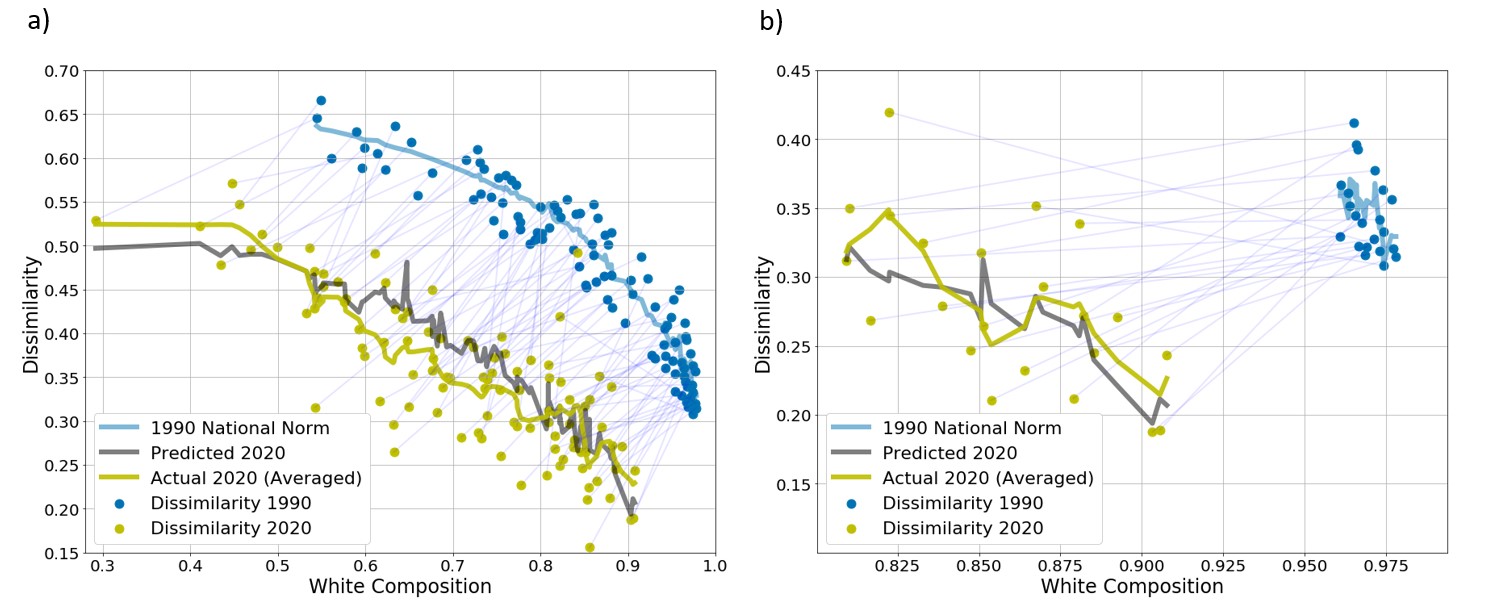}
    \caption{\textbf{Predictions of dissimilarity index versus White composition:} (a) 100 regions best represented by 1990 national compositional behavior, (b) subset of those regions with more than 96\% White composition. Shown are the 1990 values (blue dots), observed 2020 values (yellow dots) observed 2020 averaged trend (yellow curve), 1990 national norm expectations (blue curve), predicted 2020 values (grey curve). The 2020 predictions use the compositional behavior of the rest of the country and the marginal components of each region of interest. The predictions not only are accurate, but also explain how structural segregation decreases by approximately 0.2 (vertical shift between blue and yellow curves) while traditional analysis yields only approximately half of this result due to the confounding effect of the
    decreases in the White populations. }\label{fig:predictions}
\end{figure}

To illustrate the promise of such a forecasting approach we proceed as follows. First, we select 100 metropolitan areas to test, which we choose to be those that in 1990 were well-represented by the 1990 national White/non-White compositional behavior, and then attempt to `predict' the dissimilarity indices for these regions in 2020. To focus on the reliability of the overall approach, rather than the accuracy of models for forecasting regional compositions and neighborhood sizes, we will take the actual regional compositions and neighborhood size distributions from 2020 as proxies for predictions of these quantities that could be obtained with more standard approaches. Similarly, as a proxy for the extrapolation of the 2020 compositional behavior, we use the 2020 national compositional behavior that is computed from all metropolitan areas \emph{excluding} the 100 selected regions, so as to avoid biasing the results. Figure \ref{fig:predictions}a shows that the resulting predictions exhibit remarkable agreement with the actual White/non-White dissimilarity index values for 2020. Figure~\ref{fig:predictions}b underscores this agreement by highlighting the evolution of a subset of initially very similar metropolitan areas, those with a 1990 White composition of over 96\%, that nevertheless experienced quite different changes in their White compositions and dissimilarity index values. Remarkably, the large changes in the dissimilarity index for these initially very similar regions are predicted quite well, with a Pearson correlation coefficient of 0.67 between the predicted and observed changes. These results demonstrate that segregation index values can be accurately predicted with data from other regions and suggests that true forecasting of segregation should be possible.

The results in Figure \ref{fig:predictions}a also underscore several important facts regarding the historical development of segregation between 1990 and 2020 in the U.S. First, these results indicate that if the regional compositions had stayed the same between 1990 and 2020, there would have been a decrease of around 0.2 in the White/non-White dissimilarity index in all of these regions, largely independent of their 1990 composition. This is significantly larger, however, than the average observed decrease of around 0.12 that is obtained from traditional computation of dissimilarity indices. The resolution of this apparent discrepancy is that nearly all of these 100 regions simultaneously experienced a decrease in their regional White composition, which, despite the traditional view that the dissimilarity index is compositionally invariant, is strongly correlated with an increase in the dissimilarity index. Thus, working with historical dissimilarity indices while failing to adjust for changes in the regional composition can result in an underestimation of the reduction in structural segregation by roughly a factor of 2.

Despite the promise for the approach demonstrated here, clearly further work is needed before true forecasts can be made. In particular, a reliable method for extrapolating compositional behaviors will need to be developed. We also note that segregation functions can be incorporated into entirely different approaches that allow for neighborhood-level forecasts, as shown using the analogous DFFT approach of Kinkhabwala et al.  (2021\nocite{kinkhabwala2021forecasting}).

\section{Summary, Conclusions, and Future Work}
This work extends the use of information theory in segregation analyses to addresses some of the shortcomings of standard approaches through the use of segregation models.  After a brief overview of traditional segregation indices, we argued that indices tend to include assumptions that go beyond a chosen set of \emph{desirable properties} (section~\ref{sec:discord}) and noted that some properties themselves are contentious (section~\ref{sec:comp-inv}). First, we showed how information theory is intimately tied to entropic indices (section~\ref{sec:ent-ind-info}) and that information theory reveals an underlying probability model which assumes a complete lack of subgroup-dependent interactions (section~\ref{sec:logliklihood}). Leveraging this insight, we then showed how to develop underlying probability models that actually include knowledge of subgroup-dependent interactions as inferred directly from data (section~\ref{sec:Hmodel}). This inclusion led us to uncover a compositionally invariant aspect of the data that satisfied the aspatial and scale-invariant properties, the \emph{compositional behavior} that we quantified with the frustration function $f\{\pi\}$, which we then used to develop the \emph{standardized compositional distribution} (SCD) 
(section~\ref{sec:CD-xforms}). 

With the above tools in place we explored a variety of applications. We accounted for marginal differences between regions, most importantly differences in the regional composition, and established unambiguously that White/non-White structural segregation in New York City and Chicago is nearly identical while being distinct from structural segregation in Los Angeles (section~\ref{sec:SCDs}). Employing our methods to historical data, we showed that SCDs provide a detailed view of changes in structural segregation and that, following the removal of marginal differences, there have been reductions in White/non-White, White/Black, White/Hispanic, and White/Other structural segregation in the U.S. for every decade since 1990, with virtually identical trends obtained using the SCD dissimilarity and SCD isolation indices (section~\ref{sec:Historical}). Next, turning to geographical trends, we showed how to use counterfactual index values based on a national compositional behavior to identify regions of high and low structural segregation, and demonstrated that doing so uncovers trends in segregation that are largely geographically coherent and consistent with historical societal differences (section~\ref{sec:IXB}). We further showed how to group regions by their detailed compositional behavior without reduction to traditional indices, which revealed a peculiar form of intermediate White/non-White structural segregation characterized by the level of diversity in the non-White population 
(section~\ref{sec:indexless}). 
Finally, we explored the possibility of using our new concepts to develop a forecasting approach, and we discovered that failure to account for significant changes in the regional compositions could result in the dissimilarity index underestimating historical reductions in structural segregation by roughly a factor of 2 (section~\ref{sec:forecast}). 

\begin{figure}[hbt!]
    \centering
    \includegraphics[scale = 0.6]{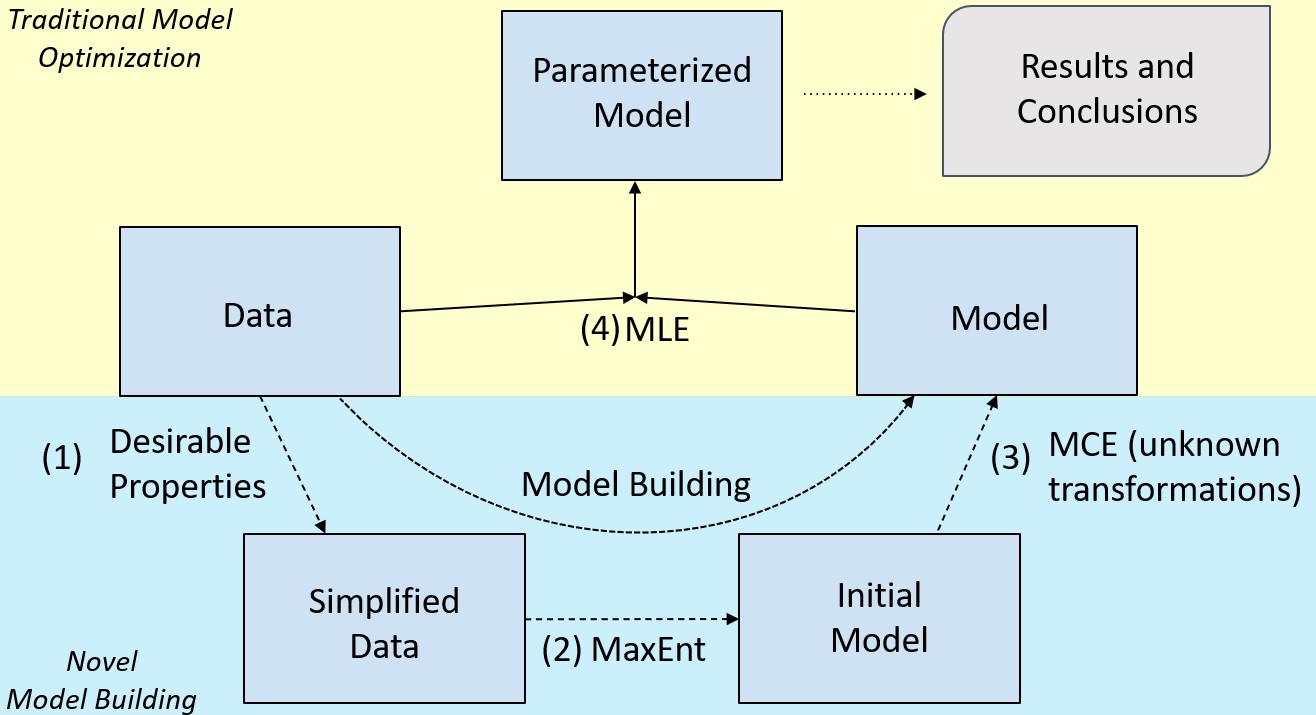}
    \caption{\textbf{Information-theory based model building}: (1) desirable properties with known transformations systematically simplify the data, (2) maximum entropy (Maxent) determines an `initial model', (3) minimum cross-entropy (MCE) allows for incorporation of properties with unkown data transformations, (4) maximum-likelihood estimation (MLE) is used to obtain a `parameterized model' from which conclusions can be drawn.}\label{framework}
\end{figure}

This work develops a systematic information-theory based approach for the processing of demographic data (Figure~\ref{framework}). Beginning with the raw \emph{data}, this approach applies (1) \emph{properties} with known simplifications to produce \emph{simplified data}. In this work, these properties included aspatiality, neighborhood ordering irrelevance, size invariance and organizational equivalence, which we used to reduce raw census population counts to compositional distributions. Next, our information-theory approach employs (2) the \emph{principle of maximum entropy} to produce an \emph{initial model} of the underlying joint probability distribution associated with the \emph{simplified data}. In this work, we showed that simplification of data down to just marginal variables resulted in a simple non-interacting model, whereas using the compositional distribution corresponded to an underlying distribution model featuring a segregation function, the `headache' $H\{\pi\}$, that is cognizant of subgroup-dependent interactions. Next, our information theory approach uses (3) the \emph{principle of minimum cross-entropy} to construct a final \emph{model} that incorporates the least-biased transformations for \emph{desirable properties} with otherwise unknown simplifications. In this work, we found that accounting of changes in regional compositions resulted in linear transformations of the headache function, characterized by the $v_m$ coefficients and the frustration function $f\{\pi\}$, that allowed us to make unbiased predictions of how compositional distributions transform under such changes. Finally, our information-theory approach uses (4) \emph{maximum likelihood estimation} to determine the model parameters from the \emph{data} to produce the final \emph{parameterized model} from which we draw all of the \emph{results and conclusions} summarized in the previous paragraph.

Looking forward, we note that the above paradigm is quite general and not limited to the specific properties employed in this work or, for that matter, to the study of segregation. A particularly promising area for extending the present work would be to start from \textit{spatial} data and to follow the same general procedure presented in Figure~\ref{framework} to produce models that appropriately incorporate the spatial dimensions of segregation. More directly, we note that we developed our models in full multigroup generality, but have presented applications only for binary comparisons for simplicity of presentation in this initial work. There is no fundamental barrier to investigation of multigroup compositional behaviors, and the resulting SCDs will appear in forthcoming publications. Finally, future applications of the approach developed here could consider binary comparisons of interest more systematically and test hypotheses that would benefit from the detailed view of structural segregation that we have demonstrated now to be possible.

\section{Acknowledgements}
\begin{small}
Y.A.K. was supported in part by funding from the National Science Foundation Graduate Research Fellowship Award
(DGE-1650441). I.C. and Y.A.K. were also supported in part by funding from (NINDS, 1R01NS116595)
and Army Research Office (ARO W911NF-18-1-0032). B.B. was supported in part by the Natural Sciences
and Engineering Research Council (NSERC PGS D).
\end{small}

\printbibliography

\clearpage
\renewcommand{\arraystretch}{1.5}
\begin{tabular}{p{5.2cm} p{5.2cm} p{5.2cm}}
 \multicolumn{3}{c}{\textbf{Table 1:} Binary indices from the compositional distribution, $P(\pi) = \frac{1}{T} \sum_{i} t_i \delta(\pi,\pi_{im})$} \\
 \hline
 & Traditional Measure& Compositional Distribution \\
 \hline
 General Form   & $\frac{1}{T} \sum_{i} t_i \ g(\pi_{im})$   &$\  \sum_{\pi} P(\pi) g(\pi)$\\
 Average Composition ($\pi_m$)&   $\frac{1}{T} \sum_i t_i \pi_{im}$ & $\sum_{\pi} P(\pi) \pi$ \\
 Dissimilarity Index&   $\frac{1}{2}\frac{1}{T \pi_m(1-\pi_m)} \sum_i  t_i |\pi_{im} - \pi_{m}|$  & $\frac{1}{2} \frac{1}{\pi_m(1-\pi_m)}\sum_{\pi} P(\pi)|\pi- \pi_{m}|$ \\
  Interaction Index   & $\frac{1}{T \pi_m} \sum_i t_i \pi_{im} (1-\pi_{im})$ & $ \frac{1}{\pi_m}\sum_\pi P(\pi) \pi (1-\pi)$\\
   Entropy Index (Binary)   & $\frac{1}{T E } \sum_i t_i \Big(  \pi_{im} \ln(\frac{\pi_{im}}{\pi_{m}})$ & $ \frac{1}{E}\sum_\pi P(\pi)\Big(  \pi \ln(\frac{\pi}{\pi_m}) $ \\ 
   & $ \quad  +(1-\pi_{im})\ln(\frac{1-\pi_{im}}{1-\pi_{m}}) \Big)$ & $ \quad  +(1-\pi)\ln(\frac{1-\pi}{1-\pi_m}) \Big)$\\
 \hline
\end{tabular}
Note that the normalization $E=-\sum_m \pi_m \ln \pi_m$ for the entropy index was not written explicitly in the table for brevity.

\clearpage

\clearpage
{\centering
\renewcommand{\arraystretch}{1.5}
\begin{tabular}{p{4cm} p{4cm} p{3cm} p{3cm} p{1.8cm}}
 \multicolumn{5}{c}{\textbf{Table 2:} Indices in major U.S. metropolitan areas following the removal of margin dependencies}\\
 \hline
 & & Original CD & Expectation \quad \quad \quad for 50\% White & SCD\\
 \hline
& New York City & 43.3\%  & 50\% & 50\%\\
 White Composition  & Los Angeles & 28.5\% & 50\% & 50\%\\
  & Chicago& 50.2\% & 50\% & 50\%\\
  \hline 
 & New York City & 0.561 & 0.528 & 0.534\\
 Dissimilarity Index  & Los Angeles & 0.523 & 0.408 & 0.432\\
 & Chicago& 0.515 & 0.516 & 0.521\\
  \hline 
  & New York City & 0.732 & 0.679 & 0.682\\
 Isolation Index  & Los Angeles & 0.798 & 0.617 &0.628 \\
  & Chicago& 0.672 & 0.673 & 0.676\\
  
 \hline
\end{tabular}}
The `expectation for 50\% White' column is obtained by incorporating the observed neighborhood sizes to create compositional distributions that result with the regional compositions being 50\% White. The standardized compositional distribution (SCD) is the expected compositional distribution \emph{if} all regions are 50\% White overall \emph{and} all neighborhoods are 1,000 in size. The isolation index is calculated for the non-White group. Neighborhood sizes are found to have a minor effect on differences between regions.

\clearpage

\clearpage

\clearpage

\clearpage

\clearpage

\clearpage

\clearpage

\clearpage

\clearpage

\clearpage

\clearpage
\appendix
\section{Information Theory Supplemental}\label{info theory}

The central premise in information theory is that the quantification of `uncertainty' has a unique form, known as \emph{entropy}. To provide an intuition for entropy, suppose we know the probabilities of various outcomes, $\{P\} = \{P_1, P_2, P_3, ...\}$, and want to quantify the amount of `choice' or `uncertainty' that these outcome probabilities imply. Mathematically, as outlined by Shannon (1948\nocite{shannon1948mathematical}), such a measure of uncertainty `$S$' should be a function that satisfies the following three conditions.

\begin{enumerate}
    \item If all outcomes are equally likely, then $S$ increases with the number of possible outcomes. 
    \item Various ways to arrive at the same outcome probabilities leaves $S$ unchanged. 
    
    \item $S$ should be a continuous function in the outcome probabilities.
\end{enumerate}

\begin{figure}[hbt!]
    \centering
    \includegraphics[scale = 0.5
    ]{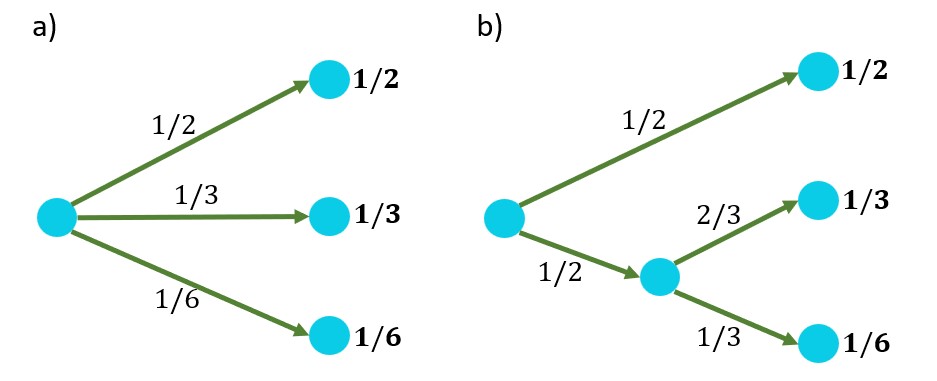}
    \caption{\textbf{Equivalent probability arrangements for condition (2)}: There are many ways to obtain a particular set of probabilities, in this case $P = \big( \frac{1}{2}, \frac{1}{3}, \frac{1}{6}\big)$. One option is to directly have a choice between these outcomes, option (a), another is to make multiple successive choices which result in these outcomes, option (b). This means that $P = \big( \frac{1}{2}, \frac{1}{3}, \frac{1}{6} \big)$ can be represented by a choice of $\big(\frac{1}{2}, \frac{1}{2} \big)$ and then, half the time, a choice of $\big( \frac{2}{3}, \frac{1}{3} \big)$. For these options to be equivalent, entropy must obey $S\big(\frac{1}{2}, \frac{1}{3}, \frac{1}{6} \big) = S \big(\frac{1}{2}, \frac{1}{2} \big) + \frac{1}{2}S \big(\frac{2}{3}, \frac{1}{3} \big)$.} \label{snapshots}
\end{figure}

The generality of these conditions cannot be overstated, the first two conditions must be satisfied for $S$ to be a measure of uncertainty on which various observers agree, while the last condition simply states that it is indeed a well-behaved function. Remarkably, the only function which satisfies these conditions is the entropy (\cite{shannon1948mathematical}), which is given by
\begin{equation}
\begin{aligned}
    S = -K \sum_k P_k \ln P_k,
\end{aligned}
\end{equation}
where $K$ is a positive constant corresponding to a unit of measurement. (We use $K = 1$.) Given that entropy is agnostic to what the outcomes represent, entropy can be considered more fundamental than common measures, such as average or variance, which would require additional quantification (\emph{i.e.}, 20\% of the time \emph{x = 1}, 36\% of the time \emph{x = 3}, \emph{etc}.). Importantly, as mentioned in the main text, entropy remains unaffected when including `potential' outcomes that never actually occur, $P_k \ln P_k = 0$ if $P_k = 0$, and, as should be expected by representing the `quantification of uncertainty', entropy is strictly non-negative.

\subsection{Principle of maximum entropy} \label{ref:PME}

Entropy has far-reaching implications. For example, consider the case of being presented with multiple options (Choice A, B and C) with no knowledge of what the outcomes represent.  It is then natural to assign each of the choices an equal probability. Historically, this equal probability assignment has been called the ``principle of insufficient reason'' or perhaps some form of Occam's razor. However, it can also be considered the most trivial case of the \emph{principle of maximum entropy}. The idea is that when presented with insufficient knowledge to make a conclusive choice, the outcomes should be assigned probabilities that satisfy the available knowledge but otherwise retain maximum uncertainty, as quantified by entropy. This allows for the systematic determination of outcome probabilities by reducing uncertainty precisely as much as a particular state of knowledge justifies.

In practice, `knowledge' can be introduced using Lagrange multipliers to incorporate constraints into an optimization problem,  resulting in finding the $\{P\}$ which maximize $S$ under the constraints resulting from the state of knowledge. For various types of knowledge (none; average; average and variance) these equations can be written
\begin{equation*}
\begin{aligned}
    \text{no knowledge:} \quad S = & - \sum_k P_k \ln P_k - \lambda \Big(\sum_k P_k - 1 \Big) \\
    \text{average:} \quad S = & -\sum_k P_k \ln P_k - \lambda \Big(\sum_k P_k - 1 \Big) - \lambda_1 \Big(\sum_k x_k P_k - \langle x \rangle \Big)\\ 
    \text{avg and var:} \quad S = & -\sum_k P_k \ln P_k - \lambda \Big(\sum_k P_k - 1 \Big) - \lambda_1 \Big(\sum_k x_k P_k - \langle x \rangle \Big) - \lambda_2 \Big(\sum_k x_k^2 P_k - \langle x^2 \rangle \Big),
\end{aligned}
\end{equation*}
where the $\lambda$ constraint maintains the normalization (that the probabilities sum to 1), the $\lambda_1$ constraint maintains the correct average value $\langle x \rangle$, and the $\lambda_2$ constraint maintains the variance, $\langle x^2 \rangle  - \langle x \rangle^2 $. Setting the derivatives of these equations to zero yields the following probability distributions,
\begin{equation*}
\begin{aligned}
    \text{no knowledge:} \quad P_k = & \frac{1}{Z}  \\
    \text{average:} \quad  P_k = & \frac{1}{Z} e^{-\lambda_1 x_k} \\
    \text{average and variance:} \quad  P_k = & \frac{1}{Z} e^{-\lambda_1 x_k -\lambda_2 x_k^2},
\end{aligned}
\end{equation*}
where $Z \equiv e^\lambda$ is used for simplicity of notation and represents the normalization factor to make the probabilities sum to unity. Finally, the Lagrange multipliers are chosen so that the constraints are correctly satisfied. At this point, it should be evident that the principle of maximum entropy leads to very familiar distributions: equal outcome probabilities for no knowledge, the exponential (Boltzmann) distribution for knowledge of the average, and a normal (Gaussian) distribution for knowledge of the mean and variance. In fact, it is well-known that familiar distributions tend to all have maximum entropy derivations.

\subsection{Information content and cross-entropy}\label{CE}
Frequently, we need to determine the degree of similarity of two distributions; for example, to quantify how well expectations of a model represent observations. To accomplish this in an information theory sense, it is useful to say that if the probability of an outcome $k$ is believed to be $P_k$ then the ``information content'' associated with that outcome is the negative log-liklihood, $I_k \equiv - \ln{P_k}$. The information content of a very high probability event is appropriately low, reflecting the fact that such an event was anticipated and hence conveys little new information. Conversely, if a highly unexpected outcome occurs it is much more meaningful and is thus rightly assigned a high information content. Using this formalism, the average information content associated with an event categorized by the distribution $P_k$ is precisely the entropy, $S = E[I] = -\sum_k P_k \ln{P_k}$. More generally, if the actual outcome probabilities $\{Q\}$ turn out differently from the model $\{P\}$, we would instead evaluate the average observed information content as
\begin{equation}
\begin{aligned}
    S(\{Q\}, \{P\}) = -\sum_k Q_k \ln P_k,
\end{aligned}
\end{equation}
which is known as the cross-entropy.
Finally, a relation known as the Gibbs' inequality ensures that the cross-entropy reaches its minimum as a function of $\{P\}$ when $\{P\} = \{Q\}$, for which it can be said the model distribution perfectly matches observations. 

Cross-entropy has several important uses. First, given a set of observations distributed according to $\{Q\}$, the parameters in a model probability distribution $\{P\}$ can be adjusted to make the cross-entropy as close to its minimum value as possible. This is precisely the well-known method of maximum likelihood estimation (MLE) for determining the parameters in a model distribution $\{P\}$. Second, the cross-entropy expression can be used in a somewhat different way. Specifically, given an initial model distribution $\{P\}$ and some new knowledge, such as a mean value different from that associated with $\{P\}$, the cross-entropy can be used to determine an updated model distribution $\{Q\}$. For this, the appropriate quantity to consider is the Kullback-Leibler (KL) divergence (\cite{shore1981properties}),
\begin{equation}
\begin{aligned}
    D_{KL} = S(\{Q\}, \{P\})-S\{Q\}  = \sum_k Q_k \ln \big( Q_k/P_k \big).
\end{aligned}
\end{equation}
Minimizing the KL divergence over $\{Q\}$, with new knowledge, is known as the principle of minimum cross-entropy (MCE) and is performed analogously to the principle of maximum entropy. Performing MCE can be thought of as reducing the expected information content of each event only as much as necessary to satisfy the new knowledge. 

\section{Non-Interacting Model of Segregation from Maximum Entropy}\label{non-int MaxEnt}

To determine the segregation model that results from a particular state of knowledge, it is possible to determine the maximum entropy distribution that results from corresponding constraints, as discussed above in Appendix \ref{ref:PME}. To apply this in our demographic context, let $p_k$ represent the probability of obtaining a particular configuration or sequence (\textit{i.e.} person 1 is White, person 2 is Black, \textit{etc.}) for a neighborhood and suppose that the `available knowledge' is the regional composition and neighborhood size $t_i$. The maximum entropy distribution is one which satisfies
\begin{equation}
\begin{aligned}
    0 = \frac{\partial}{\partial p_k} \Big(-\sum_kp_k \ln p_k - \lambda \big(\sum_k p_k - 1 \big) - \sum_m \lambda_m(\sum_{k} t_{km} p_k - t\pi_m) \Big) \quad \forall p_k,
\end{aligned}
\end{equation}
where $t_{km}$ is the number of $m$ individuals in a particular sequence $k$ and the various $\lambda$ represent Lagrange multipliers. The first term is the entropy, the second term enforces the probability normalization constraint, and the third term enforces the constraint that for each subgroup the expected composition matches the actual  regional composition, $t\pi_m$. Solution of the above equation leads to a probability form of
\begin{equation}
\begin{aligned}
   p_k = \frac{1}{Z}e^{-\sum_m \lambda_m t_{km}}.
\end{aligned}
\end{equation}

The occupancy distribution, indicating the number of individuals of each subgroup regardless of the sequence, is given by summing all sequences that leave the occupancy unchanged
\begin{equation}
\begin{aligned}
   P\{t\} = \sum_{k\in \{t\}}p_k = \sum_{k \in \{t\}} \frac{1}{Z}e^{-\sum_m \lambda_m t_{km}} =  \frac{1}{Z} \frac{t_i!}{\prod_m t_{im}!} e^{-\sum_m \lambda_m t_{im}},
\end{aligned}
\end{equation}
where $t_{im}$ is the neighborhood occupancy of subgroup $m$. Solving for the Lagrange multipliers to yield the correct regional composition ultimately gives
\begin{equation}
\begin{aligned}
    P\{t\}= \frac{t_i!}{\prod_m t_{im}!} \prod_m \pi_m^{t_{im}},
\end{aligned}
\end{equation}
precisely the multinomial distribution.

\section{Multinomial Distribution and Entropy-based Indices}\label{multinomial and ent}

The rightmost product of the multinomial distribution can be written in an exponential form,
\begin{equation}
\begin{aligned}
    \prod_m \pi_m^{t_{im}} = e^{\sum_m t_{im} \ln (\pi_m)} =   e^{t_i \sum_m \pi_{im} \ln (\pi_m)}.
\end{aligned} \label{latter product}
\end{equation}
As for the combinatorial prefactor, we will use Sterling's approximation,
\begin{equation}
\begin{aligned}
\ln{(N!)} \approx N\ln{N}-N,
\end{aligned} 
\end{equation}
which removes discreteness associated with finite neighborhood size by presuming that the neighborhoods are large ($\gtrsim$ 100). 
For the combinatorial term, this expansion gives
\begin{equation}
\begin{aligned}
\ln \Bigg( \frac{t_i!}{\prod_m t_{im}!} \Bigg) \approx & \ t_i \ln t_i - t_i -\sum_m \big(t_{im}\ln t_{im} - t_{im}\big) \\
  = & -t_i \sum_m \pi_{im} \ln \pi_{im}, \\
\end{aligned} \label{comb_result}
\end{equation}
where we have used $\sum_m t_{im} = t_i$ and $\pi_{im} = t_{im}/t_i$.
Combining equation \ref{latter product} and \ref{comb_result} leads to the overall log-probability for the multinomial distribution of
\begin{equation}
\begin{aligned}
    \ln P\{t\} = & \  \ln \Bigg( \frac{t_i!}{\prod_m t_{im}!} \Bigg) + \ln \Bigg(\prod_m \pi_m^{t_{im}}  \Bigg)\\
    \approx & -t_i \sum_m \pi_{im} \ln \Big( \frac{\pi_{im}}{\pi_{m}} \Big) .
\end{aligned}
\end{equation}
Therefore, as described in the main text, we can relate the neighborhood divergence index and the entropy index for the region directly to $\ln P\{t\}$ through
\begin{equation}
\begin{aligned}
    D_i & \equiv \sum_m \pi_{im} \ln \Big( \frac{\pi_{im}}{\pi_{m}} \Big) \approx -\frac{1}{t_i} \ln P\{t_i\}\\ 
    S & \equiv  \frac{1}{TE}\sum_i t_i \sum_m \pi_{im} \ln \Big( \frac{\pi_{im}}{\pi_{m}} \Big) \approx -\frac{1}{TE} \sum_i \ln P\{t_i\},
\end{aligned}
\end{equation}
where $D_i$ is the neighborhood divergence index, $S$ is the entropy index for the region, $T$ is the total regional population and  $E = -\sum_m \pi_{m} \ln \pi_{m}$. Although this derivation used Sterling's approximation, for most practical purposes it can be considered exact, as we will discuss in more detail in forthcoming work.

\section{Segregation Functions from Information Theory}

\subsection{Interacting Model of Segregation from Maximum Entropy}\label{App:Int1}

Suppose we have a compositional distribution $P\{\pi\}$ and the size distribution of neighborhoods in a region $P(t)$. The maximum entropy equation is then
\small
\begin{equation}
\begin{aligned}
    0 = \frac{\partial}{\partial p_{kt}} \Big(\sum_{k,\ t} - p_{kt} \ln p_{kt} - \lambda \big(\sum_{k,\ t} p_{kt} - 1 \big) -  \sum_{\{\pi\}} \lambda{\{\pi\}} \big( \sum_{k\in \{\pi\}} tp_{kt} - T P\{\pi\}\big) -\sum_{t} \lambda(t) \big(\sum_k p_{kt} - P(t) \big)\Big)
\end{aligned}
\end{equation}
\normalsize
for all $P_{kt}$, where $p_{kt}$ represents the probability of a neighborhood having a particular `sequence' or `configuration' $k$ and a size $t$, $T P\{\pi\}$ is the observed number of individuals being in neighborhoods of composition $\{\pi\}$ (which is made to match a size-weighted sum of model distributions), and the constraints are introduced as Lagrange multipliers. The solution of the above equation is a distribution of the form
\begin{equation}
\begin{aligned}
   p_{kt} = \frac{1}{Z}e^{-t \lambda{\{\pi\} - \lambda(t)}},
\end{aligned}\label{eq:pkt}
\end{equation}
so that probability of obtaining a neighborhood of composition $\{\pi\}$ and size $t$ is
\begin{equation}
\begin{aligned}
   p_{\{ \pi \} t} = \sum_{k \in \{\pi\}} \frac{1}{Z}e^{-t \lambda{\{\pi\} - \lambda(t)}} = \frac{1}{Z} \frac{t!}{\prod_m (t\pi_{im})!} e^{-t \lambda{\{\pi\} - \lambda(t)}}.
\end{aligned}
\end{equation}
For a neighborhood with \emph{known} size $t_i$, this becomes
\begin{equation}
\begin{aligned}
   p_{t_i}\{ \pi \} = \frac{p_{\{ \pi \} t_i}}{\sum_{\{\pi\}} p_{\{ \pi \} t_i}} = \frac{1}{Z_{t_i}} \frac{t_i!}{\prod_m (t_i\pi_{im})!} e^{-t_i \lambda{\{\pi\}}},
\end{aligned}
\end{equation}
where $Z_{t_i}\equiv\sum_{\{\pi\}} \frac{t_i!}{\prod_m (t_i\pi_{im})!} e^{-t_i \lambda{\{\pi\}}}$ is the appropriate normalization for size $t_i$. Renaming $\lambda \{\pi \}$ as $H\{\pi \}$ and using the fact that $t_{im} = t_i\pi_{im}$ then yields equation \ref{dfft form} from the main text,
\begin{equation}
\begin{aligned}
    P\{t\}= \frac{1}{Z_{t_i}} \frac{t_i!}{\prod_m t_{im}!} e^{-t H\{\pi\} }.
\end{aligned}
\end{equation}

\subsection{Accounting for Compositional Variance with Minimum Cross Entropy}\label{CVMCE}
Suppose that after solving for $p_{kt}$ in equation \ref{eq:pkt}, we want to update the distribution with knowledge of a different regional composition. Let the new probability distribution be given by $q_{kt}$, which makes the cross-entropy (MCE) equation
\begin{equation}
\begin{aligned}
    0 = \frac{\partial}{\partial q_{kt}} \Big(\sum_{k,\ t} q_{kt} \ln \big(q_{kt}/p_{kt}\big)  - \lambda \big(\sum_{k,\ t} q_{kt} - 1 \big) -\sum_m \lambda_m(\sum_{k} t_{km} q_{kt} - t\pi_m) \Big) \quad \forall q_{kt},
\end{aligned}
\end{equation}
where the first term is the cross-entropy to be minimized and the remaining terms are the Lagrange multipliers maintaining normalization and the new regional composition.
The solution of this new equation then gives
\begin{equation}
\begin{aligned}
   q_{kt} = \frac{p_{kt}}{Z}e^{-\sum_m \lambda_m t_{km}} \quad \mbox{and}  \quad q_{\{\pi\}t} = \frac{p_{\{\pi\}t}}{Z}e^{-t\sum_m \lambda_m \pi_{im}},
\end{aligned}
\end{equation}
which means that, for a neighborhood of known size $t_i$, the distribution is
\begin{equation}
\begin{aligned}
   q_{t_i}\{ \pi \} = \frac{q_{\{ \pi \} t_i}}{\sum_{\{\pi\}} q_{\{ \pi \} t_i}} = \frac{1}{Z_{t_i}} \frac{t_i!}{\prod_m (t_i \pi_{im})!} e^{-t_i (\lambda{\{\pi\}+\sum_m \lambda_m \pi_{im})}},
\end{aligned}
\end{equation}
where $Z_{t_i} \equiv \sum_{\{\pi\}} \frac{t_i!}{\prod_m (t_i\pi_{im})!} e^{-t_i (\lambda{\{\pi\}+\sum_m \lambda_m \pi_{im})}}$ is the appropriate normalization for size $t_i$. Renaming $\lambda \{\pi \}$ as $f\{\pi \}$ and $\lambda_m$ as $v_m$ then yields equation \ref{dfft form2} from the main text and demonstrates that compositional invariance results in partitioning of $H\{\pi\}$ into
\begin{equation}
\begin{aligned}
    H \{\pi\} = \sum_m \pi_{im} v_m + f\{\pi\},
\end{aligned}
\end{equation}
where $v_m$ accounts for differences in the regional composition and $f\{\pi\}$ is compositionally invariant.

\section{(Binary) Numerical Optimization of Parameters}\label{num_op}
In this manuscript we have obtained numerical results for binary segregation (\emph{e.g.,} White/non-White) using the models from equation \ref{dfft form} and \ref{dfft form binary} from the main text. These models were obtained using the principle of maximum entropy (Maxent) and the principle of minimum cross-entropy (MCE). Here, we will first demonstrate how to optimize the $H(\pi)$ function (the binary case of $H \{\pi\}$) with data from a single region. Then, we will show how to perform an optimization for multiple regions using a unique $f(\pi)$ with different values of $v_m$ for each region. 

\subsection{Single Region Optimization} \label{Opt1}
To determine the unknown function $H(\pi)$, which is used to recreate the observed compositional distribution, we employ maximum likelihood estimation (MLE). This procedure is carried out by minimizing the negative logarithm of the probability (sometimes called the negative log-likelihood) over the observed neighborhood compositions,
\begin{equation}
\begin{aligned}
    -\sum_i \ln {P(\pi_{im})} = \sum_i \bigg(t_{i} H(\pi_{im})-\ln \Big(\frac{t_i!}{t_{im}!(t_i-t_{im})!} \Big) + \ln {Z_{t_i}}\bigg).
\end{aligned}\label{MLE1}
\end{equation}
First, we that that, from a numerical perspective, it is advantageous to explicitly account for the dominant contribution of the combinatorial factor, as determined from equation~\ref{comb_result}, and combine it with $H$, defining a new variable $\Bar{H}(\pi) = H(\pi)+\pi \ln \pi+(1-\pi)\ln (1-\pi)$. This is advantageous because of the empirical observation that compositional distributions tend to be highly flat (compared to a sharply peaked binomial distribution) and so the dominant portions of $H(\pi)$ and the combinatorial term are expected to cancel, leaving the more well behaved function $\Bar{H}(\pi)$. The optimization can then be performed over $\Bar{H}(\pi)$, with equation~\ref{MLE1} becoming
\begin{equation}
\begin{aligned}
   g = \sum_i \Big(t_{i} \Bar{H}(\pi_{im}) + \ln {Z_{t_i}}\Big) + C,
\end{aligned} \label{g}
\end{equation}
where $C$ is an irrelevant constant with respect to our optimization parameters. Further, although this equation is exact, in practice we will optimize for a discretized approximation, $\Tilde{g}$, which presumes that, over sufficiently small `compositional bins' (\textit{e.g.,} cases with compositions between 46 and 48\% White), $\Bar{H}(\pi)$ stays constant. Denoting the chosen compositional bins as $\Tilde{\pi}$, the first term of equation~\ref{g} simplifies as
\begin{equation}
\begin{aligned}
   \sum_i t_{i} \Bar{H}(\pi_{im}) = T \sum_\pi P(\pi) \Bar{H}(\pi) \approx T \sum_{\Tilde{\pi}} \Bar{H}(\Tilde{\pi}) \sum_{\pi \in \Tilde{\pi}} P(\pi)  = T \sum_{\Tilde{\pi}} \Bar{H}(\Tilde{\pi}) P(\Tilde{\pi}),
\end{aligned}\label{g1}
\end{equation}
where $P(\pi)$ is the compositional distribution of a region. We then see that this discretization of $\Bar{H}(\pi)$ results in a discretization of the compositional distribution, $P(\Tilde{\pi})$, which represents the probability of finding an individual in a neighborhood with a composition in bin $\Tilde{\pi}$. We emphasize that this approximation can be made to any desired accuracy. For the applications in this work, for computational expediency, we typically employ 50 or 100 compositional bins.

Continuing with our approximation $\Tilde{g}$ of  equation~\ref{g}, the second term becomes
\begin{equation}
\begin{aligned}
   \sum_i \ln Z_{t_i} = \sum_t N(t) \ln Z_t = \sum_t N(t) \ln \Bigg(\sum_{\Tilde{\pi}} e^{-t\Bar{H}(\Tilde{\pi})} \sum_{\pi \in \Tilde{\pi}} e^{\ln \Big(\frac{t!}{(t\pi)!(t(1-\pi))!} \Big)+t\big(\pi \ln \pi +(1-\pi)\ln(1-\pi)\big)}\Bigg),
\end{aligned}
\end{equation}
where $N(t)$ counts how many times a neighborhood of size $t$ occurs in the region.
Here, the summation of $\sum_{\pi \in \Tilde{\pi}}$ is independent of both $\Bar{H}(\Tilde{\pi})$ and the observed compositional distribution and, hence, can be calculated once and represented by $u(t, \Tilde{\pi})$, resulting in
\begin{equation}
\begin{aligned}
    \Tilde{g} = T \sum_{\Tilde{\pi}} \Bar{H}(\Tilde{\pi}) P(\Tilde{\pi})  + \sum_t N(t) \ln \Big( \sum_{\Tilde{\pi}} u(t, \Tilde{\pi}) e^{-t\Bar{H}(\Tilde{\pi})}\Big) + C.
\end{aligned}\label{eq:tg}
\end{equation}

Finally, when differentiated over $\Bar{H}(\Tilde{\pi})$, equation~\ref{eq:tg} gives the gradient
\begin{equation}
\begin{aligned}
    \frac{\partial \Tilde{g}}{\partial \Bar{H}(\Tilde{\pi})} = T P(\Tilde{\pi}) - \sum_t \frac{t N(t)}{Z_t} u(t, \Tilde{\pi}) e^{-t\Bar{H}(\Tilde{\pi})},
\end{aligned}
\end{equation}
which, appropriately, gives zero when the model's prediction for the number of individuals experiencing a neighborhood composition (which falls into a compositional bin $\Tilde{\pi}$) matches observation. Access to this gradient then allows for the application of numerical optimization algorithms to efficiently compute $\Bar{H}(\Tilde{\pi})$, and thus obtain $H(\pi)$ to desired accuracy.

\subsection{Multiple Region Optimization}\label{multiregion}
The optimization procedure with multiple regions follows a similar procedure as section~\ref{Opt1}, but now with an additional optimization over regions $R$,
\begin{equation}
\begin{aligned}
    -\sum_R \sum_{i \in R} \ln {P_R(\pi_{im})} = \sum_R \sum_{i \in R} \bigg(t_{i} H_R(\pi_{im})-\ln \Big(\frac{t_i!}{t_{im}!(t_i-t_{im})!} \Big) + \ln {Z_{R,t_i}}\bigg),
\end{aligned} \label{MLE2}
\end{equation}
with $H_R(\pi) = f(\pi) + v_R \pi$, where $f(\pi)$ is region independent. Next, following the logic from the previous subsection, we let $\Bar{f}(\pi) = f(\pi)+\pi \ln \pi+(1-\pi)\ln (1-\pi)$, transforming equation \ref{MLE2} into
\begin{equation}
\begin{aligned}
   g = \sum_R \sum_{i \in R} \Big(t_{i} \Bar{f}(\pi_{im})+t_{i} v_R \pi_{im} + \ln {Z_{R,t_i}}\Big) + C.
\end{aligned} \label{G}
\end{equation}
As in the previous section, we optimize over the discretized version $\Tilde{g}$, whose first term becomes 
\begin{equation}
\begin{aligned}
   \sum_R \sum_{i \in R}  t_{i} \Bar{f}(\pi_{im}) = \sum_R T_R \sum_\pi P_R(\pi) \Bar{f}(\pi) =\sum_{\Tilde{\pi}} \Bar{f}(\Tilde{\pi}) \sum_R T_R \sum_{\pi \in \Tilde{\pi}} P_R(\pi) = \sum_{\Tilde{\pi}} \Bar{f}(\Tilde{\pi}) \sum_R T_R P_R(\Tilde{\pi}).
\end{aligned}
\end{equation}
Next, using $P(\Tilde{\pi})$ for consistency, the second term becomes
\begin{equation}
\begin{aligned}
   \sum_R \sum_{i \in R} t_{i} v_R \pi_{im} = \sum_R T_R \sum_{\Tilde{\pi}}  v_R P_R(\Tilde{\pi}) \Tilde{\pi} = \sum_R v_R T_R \pi_R
\end{aligned},
\end{equation}
where $\Tilde{\pi}$ is taken to be the average composition of bin $\Tilde{\pi}$ and $\pi_R$ is the average composition in region $R$. Finally, the third term becomes
\begin{equation}
\begin{aligned}
   \sum_R \sum_{i \in R} \ln Z_{R, t_i} = \sum_R \sum_t N_R(t) \ln Z_{R,t} = \sum_R \sum_t N_R(t) \ln \Bigg(\sum_{\Tilde{\pi}} u(t, \Tilde{\pi}) e^{-t(\Bar{f}(\Tilde{\pi}) + \Tilde{\pi} v_R) }\Bigg),
\end{aligned}
\end{equation}
and the discretized version of equation~\ref{G} is then the sum of the above terms,
\begin{equation}
\begin{aligned}
   \Tilde{g}= \sum_{\Tilde{\pi}} \Bar{f}(\Tilde{\pi}) \sum_R T_R P_R(\Tilde{\pi}) + \sum_R v_R T_R \pi_R + \sum_R \sum_t N_R(t) \ln \Bigg(\sum_{\Tilde{\pi}} u(t, \Tilde{\pi}) e^{-t(\Bar{f}(\Tilde{\pi}) + \Tilde{\pi} v_R) }\Bigg) + C.
\end{aligned} 
\end{equation}
The gradient components of this function with respect to the $\Bar{f}(\Tilde{\pi})$ variables are
\begin{equation}
\begin{aligned}
    \frac{\partial \Tilde{g}}{\partial \Bar{f}(\Tilde{\pi})} = \sum_R T_R P_R(\Tilde{\pi}) - \sum_R \sum_t \frac{t N_R(t)}{Z_{R,t}} u(t, \Tilde{\pi}) e^{-t(\Bar{f}(\Tilde{\pi}) +\Tilde{\pi}v_R)},
\end{aligned}
\end{equation}
which give zero when the model's prediction for the total number of individuals across all regions experiencing a neighborhood composition that falls into composition bin $\Tilde{\pi}$  matches observation. Finally, the gradient components with respect to the $v_R$ variables are
\begin{equation}
\begin{aligned}
    \frac{\partial \Tilde{g}}{\partial v_R} = T_R \pi_R  - \sum_t \frac{t N_R(t)}{Z_{R,t}} \sum_{\Tilde{\pi}}\Tilde{\pi} u(t, \Tilde{\pi}) e^{-t(\Bar{f}(\Tilde{\pi}) +\Tilde{\pi}v_R)},
\end{aligned}
\end{equation}
which give zero when the model's prediction for the total number of individuals of interest (\emph{e.g.,} White) in region $R$ matches observation. Finally, as above, these gradient components can then be employed with a numerical optimization algorithm to optimize $\Tilde{g}$ and thereby determine the model parameters.

\section{Supplemental Tables and Figures}

\subsection{Non-Smoothed Distributions} \label{non-smoothed}
To demonstrate how SCDs appear in the absence of Gaussian smoothing, the following represents results of section \ref{sec:SCDs} with 20 and 50 bin-discretization of the headache function. The results presented in the main text constitute a 50 bin-discretization with Gaussian smoothing. 

\begin{figure}[hbt!]
    \includegraphics[width=6.5in,keepaspectratio]{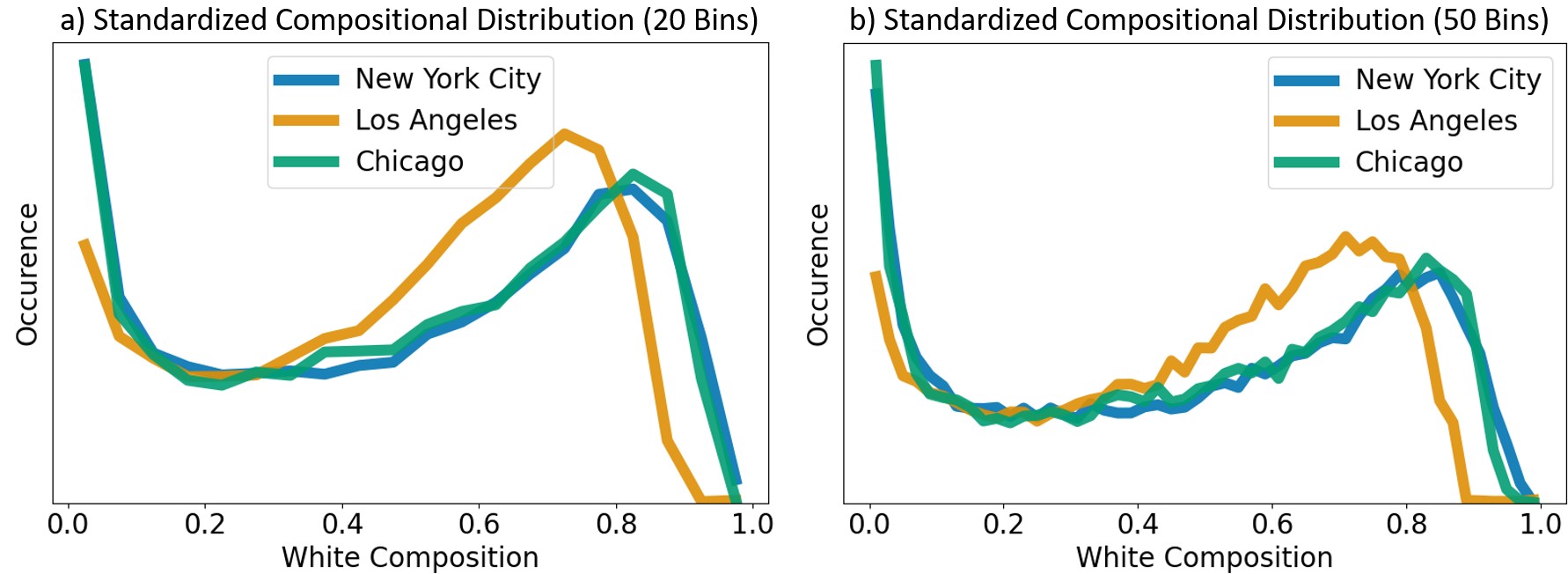}
    \caption{\textbf{White/non-White standardized compositional distributions of the three largest U.S. metropolitan areas}: headache function discretized into (a) 20 and (b) 50 bin intervals. The apparent difference in vertical range results from the high prevalence of entirely non-White neighborhoods becoming more pronounced with finer discretization. The more pronounced fluctuations in the 50-bin case are the result of statistical fluctuations from the smaller number of cases falling into each bin.}
\end{figure}

\subsection{National Predictions of Regional Segregation}\label{NatPredReg}

\subsubsection{White/Non-White}
In sections \ref{sec:IXA} and \ref{sec:IXB} index predictions using a national compositional behavior were presented. The 30 metropolitan areas best represented by the national predictions are shown in Table 3, with the least/most segregated shown in Table 4 and 5, respectively. 

\begin{figure}[hbt!]
    \includegraphics[width=6.4in]{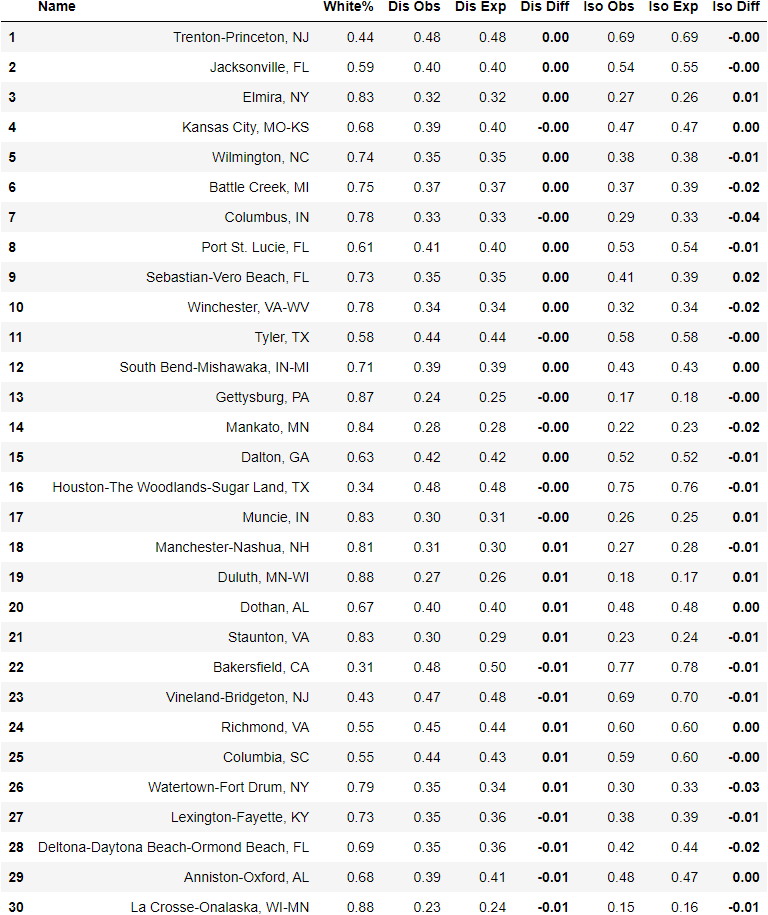}
    \textbf{Table 3}. Metropolitan areas that had their 2020 White/non-White dissimilarity index predicted well using the 2020 White/non-White national compositional behavior.
\end{figure}

\begin{figure}[hbt!]
    \includegraphics[width=6.4in]{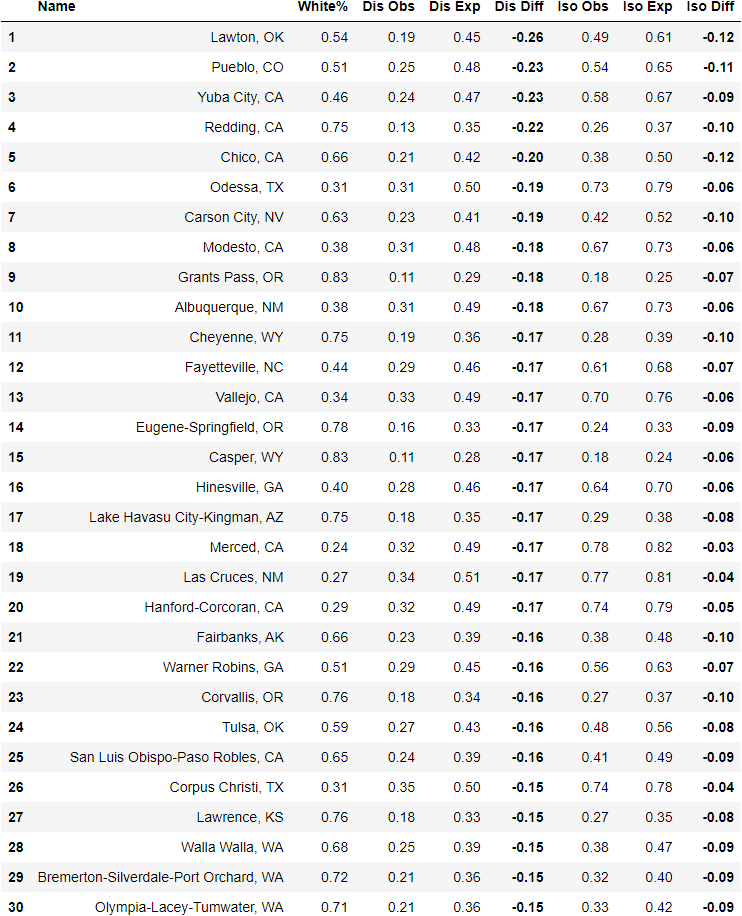}
    \textbf{Table 4}: White/non-White least segregated metropolitan areas in 2020 as determined by their dissimilarity index difference from 2020 White/non-White national expectations.
\end{figure}

\begin{figure}[hbt!]
    \includegraphics[width=6.4in]{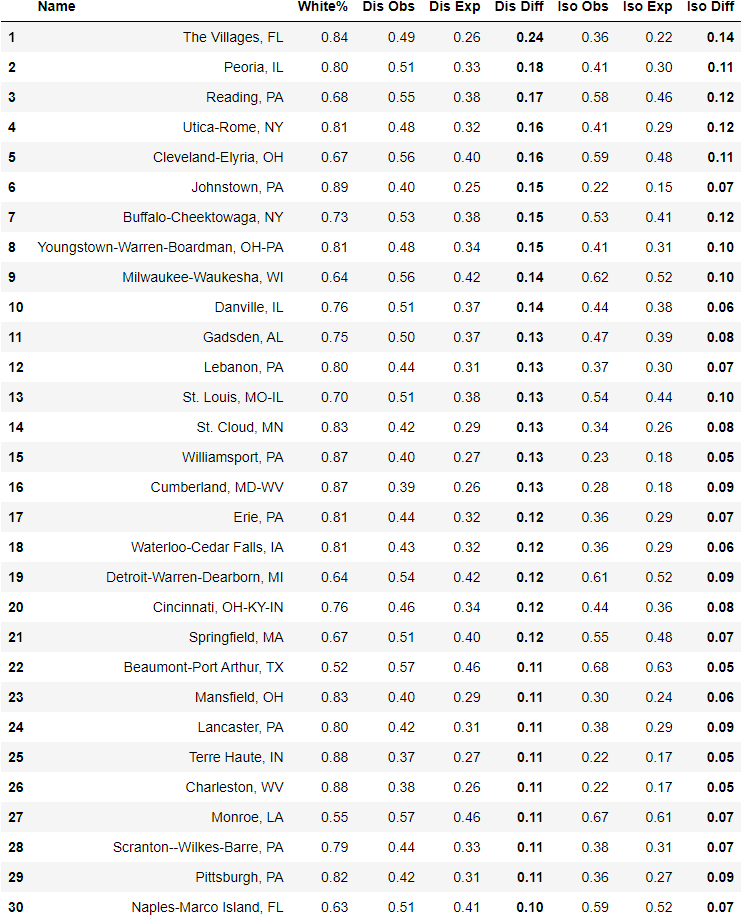}
    \textbf{Table 5}: White/non-White most segregated metropolitan areas in 2020 as determined by their dissimilarity index difference from 2020 White/non-White national expectations.
\end{figure}

\subsubsection{White/Black} \label{Ap:w/b}
The plots of \ref{sec:IXA} and \ref{sec:IXB} for White/Black are shown in Figure 14. The 30 metropolitan areas best represented by the national predictions are shown in Table 6, with the least/most segregated shown in Table 7 and 8, respectively. 

\begin{figure}[hbt!] 
    \includegraphics[width=6.5in,keepaspectratio]{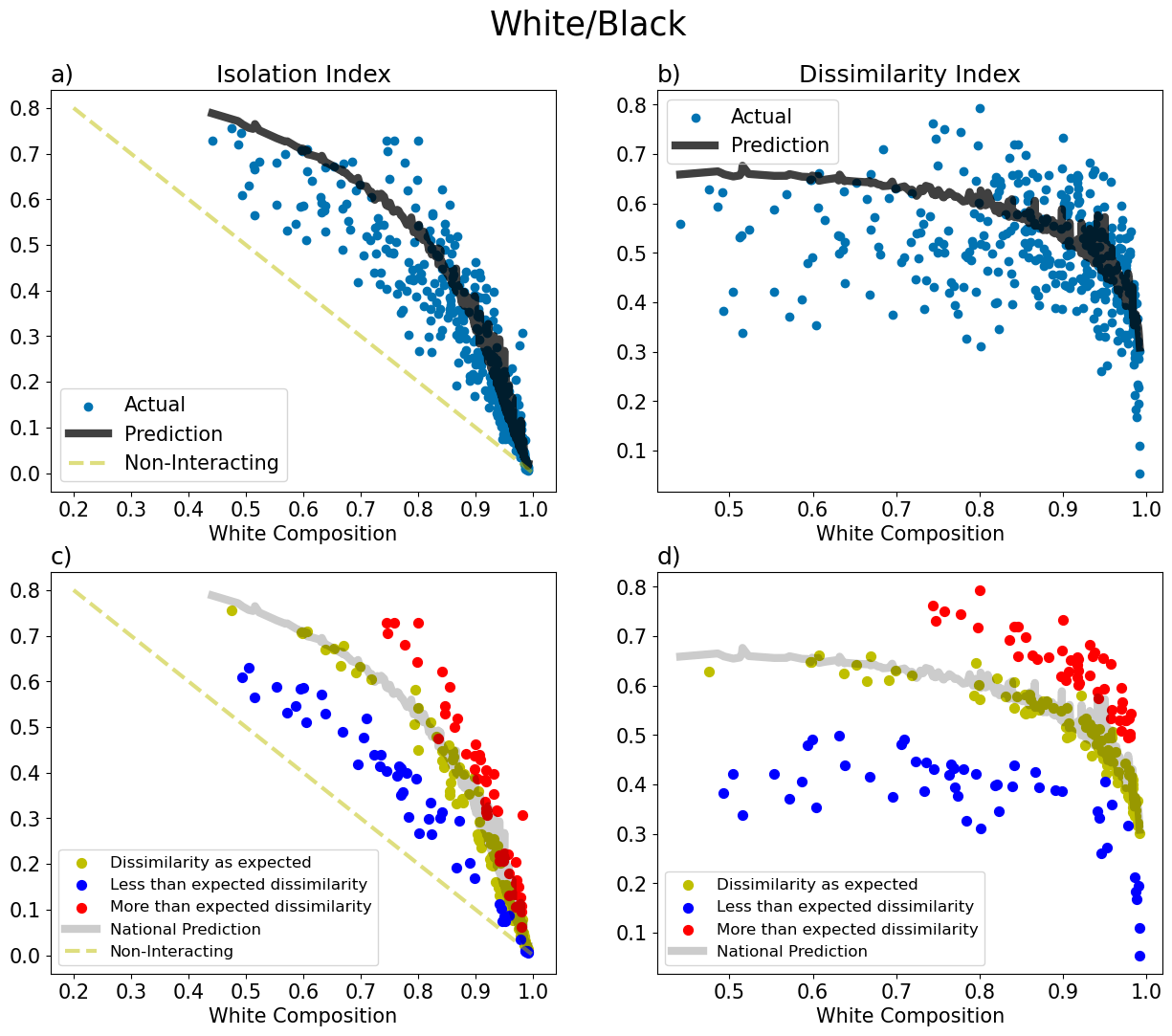}
    \caption{\textbf{Dependence of White/Black segregation indices on the regional White composition}: (a,c) isolation index, (b,d) dissimilarity index, observed data (blue dots in a,b), predictions based on 2020 White/Black national compositional behavior (points connected by grey curves), expectations from non-interacting model (dashed yellow line), 100 regions with the best prediction for the dissimilarity index (yellow dots in c,d), and the 50 regions with the most over-estimated (red dots in c,d) and under-estimated (blue dots in c,d) dissimilarity indices. Data are for neighborhoods taken to be block groups, and the data in (a-b) are for all U.S. metropolitan areas in 2020.  The isolation index values for each group of regions follow the same trends relative to the national norm as the dissimilarity index values, indicating that the differences from the norm expectations are reflective of regions of stronger or weaker structural segregation.} 
    \label{fig:White-black_compositional_dependence}
\end{figure}

\begin{figure}[hbt!]
    \includegraphics[width=6.4in]{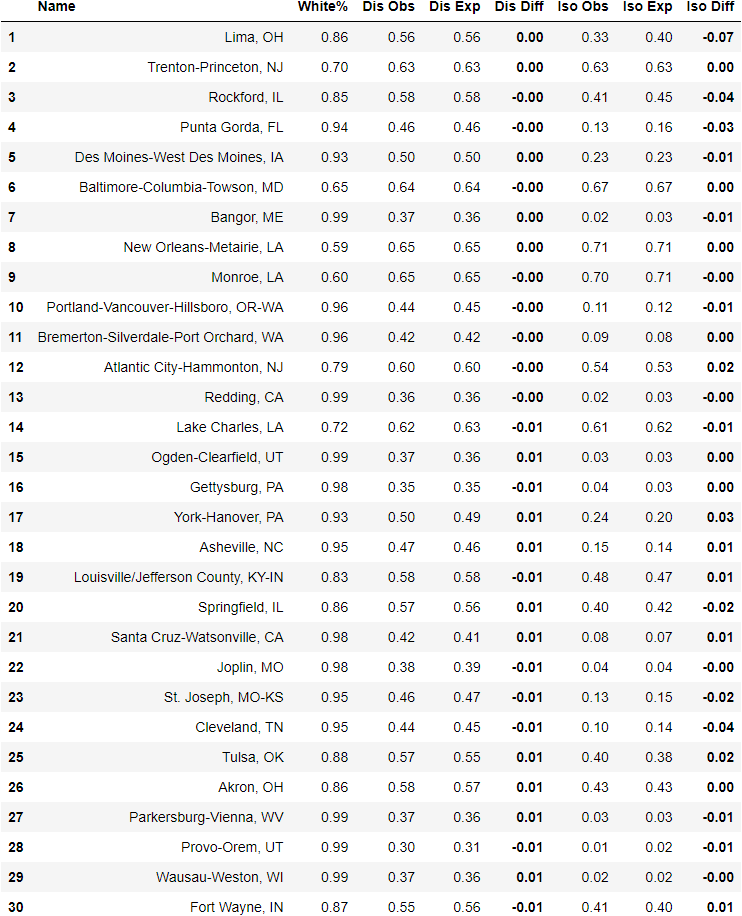}
    \textbf{Table 6}: Metropolitan areas that had their 2020 White/Black dissimilarity index predicted well using the 2020 White/Black national compositional behavior
\end{figure}

\begin{figure}[hbt!]
    \includegraphics[width=6.4in]{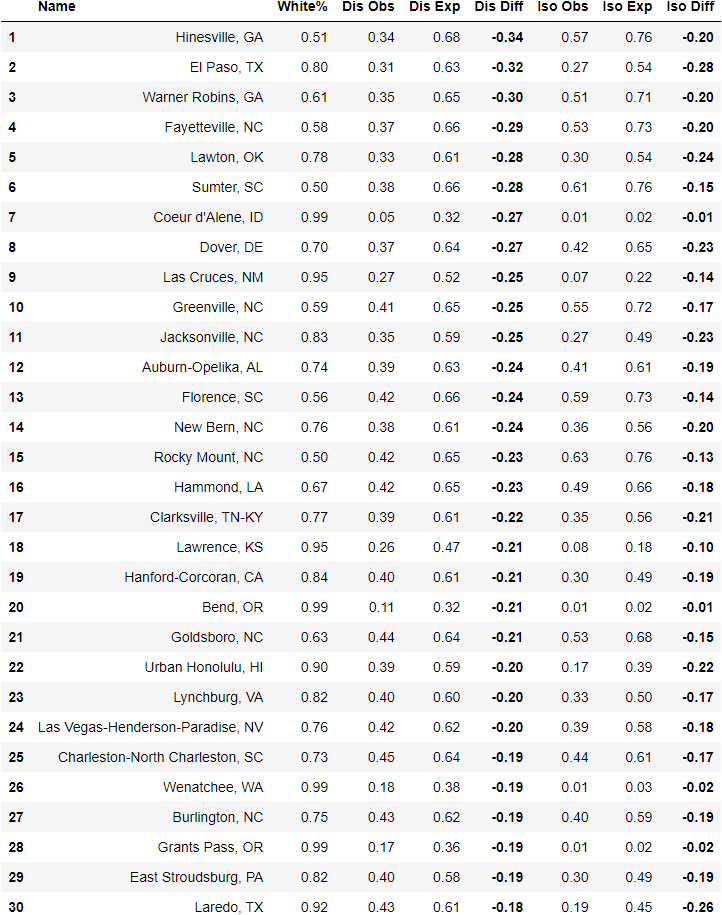}
    \textbf{Table 7}: White/Black least segregated metropolitan areas in 2020 as determined by their dissimilarity index difference from 2020 White/Black national expectations.
\end{figure}

\begin{figure}[hbt!]
    \includegraphics[width=6.4in]{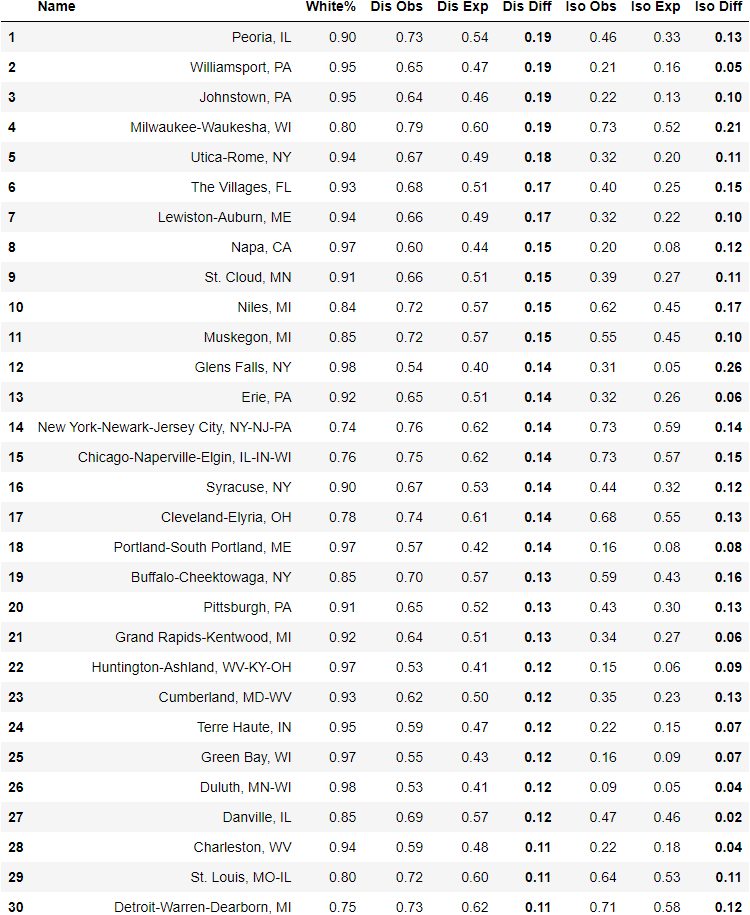}
    \textbf{Table 8}: White/Black most segregated metropolitan areas in 2020 as determined by their dissimilarity index difference from 2020 White/Black national expectations.
\end{figure}

\subsubsection{White/Hispanic}
The plots of \ref{sec:IXA} and \ref{sec:IXB} for White/Hispanic are shown in Figure 15. The 30 metropolitan areas best represented by the national predictions are shown in Table 9, with the least/most segregated shown in Table 10 and 11, respectively. 
\begin{figure}[hbt!]
    \includegraphics[width=6.5in,keepaspectratio]{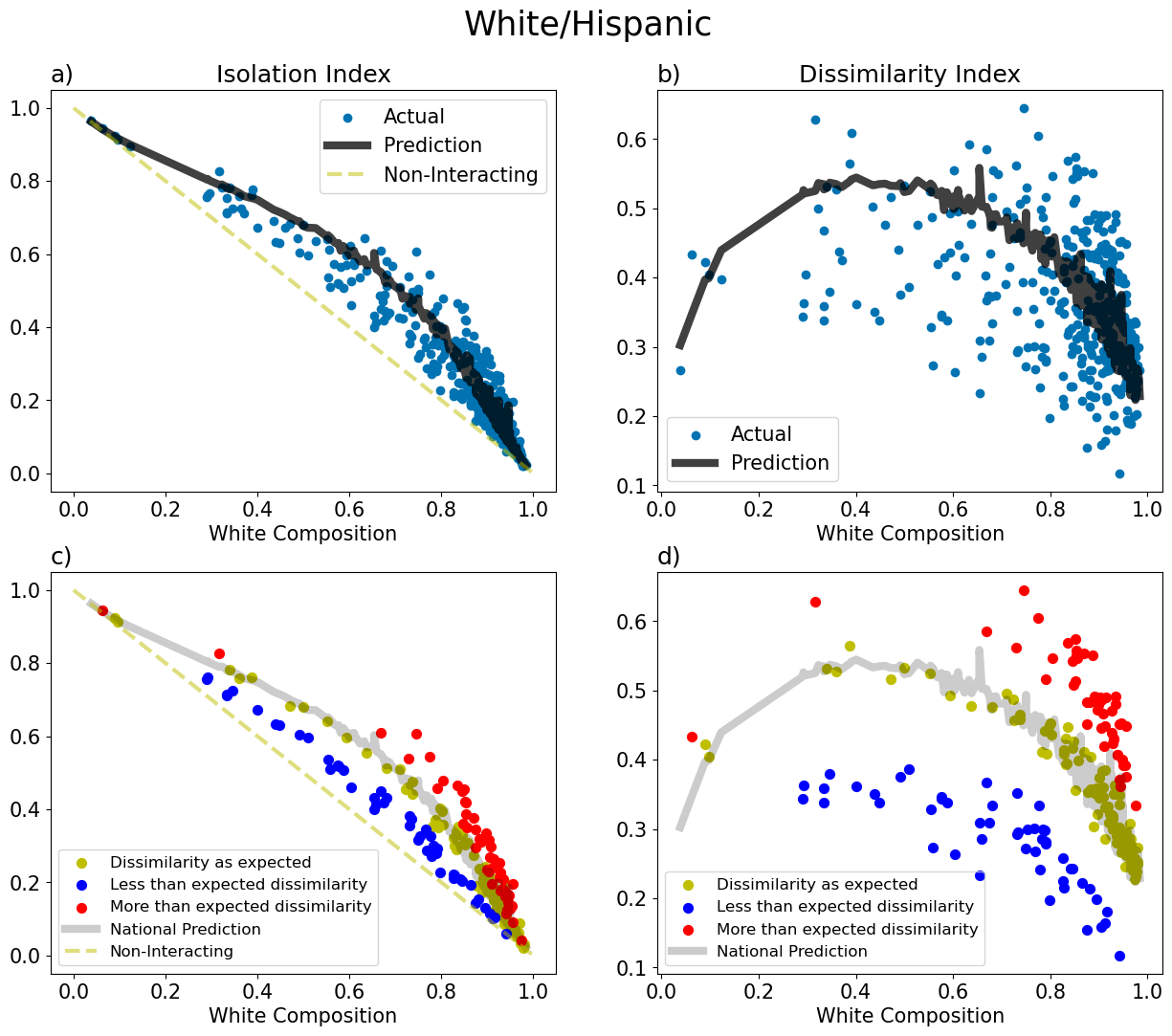}
    \caption{\textbf{Dependence of White/Hispanic segregation indices on the regional White composition}: same conventions as Figure 14, with the data again indicative that differences from the national norm consistently represent regions of stronger or weaker structural segregation.}
    \label{fig:White-hispanic compositional dependence}
\end{figure}

\begin{figure}[hbt!]
    \includegraphics[width=6.4in]{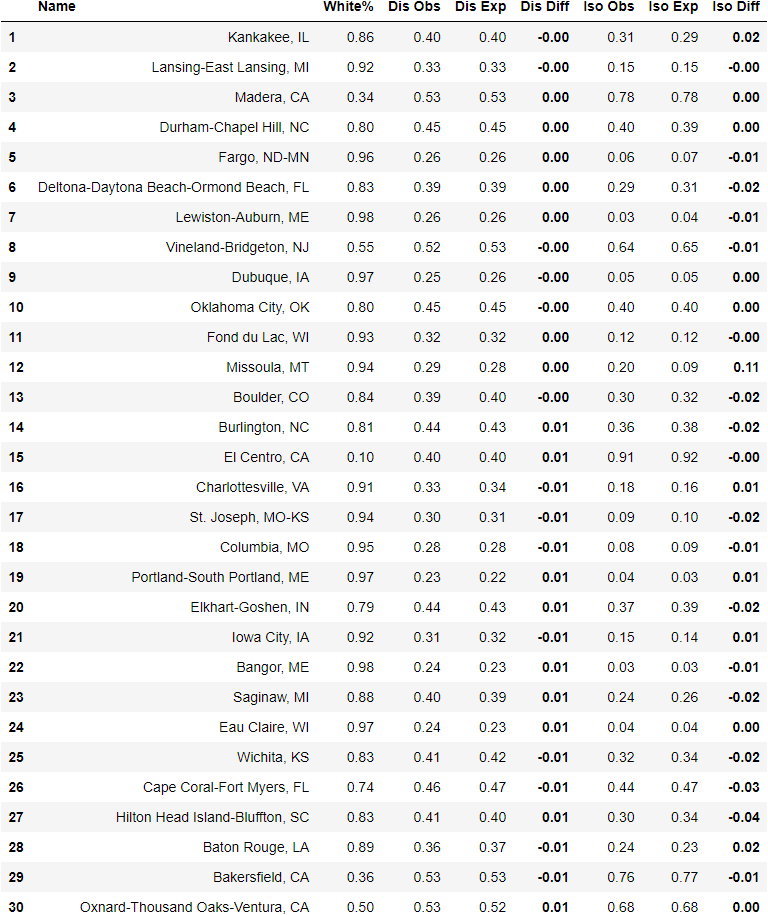}
    \textbf{Table 9}: Metropolitan areas that had their 2020 White/Hispanic dissimilarity index predicted well using the 2020 White/Hispanic national compositional behavior
\end{figure}

\begin{figure}[hbt!]
    \includegraphics[width=6.4in]{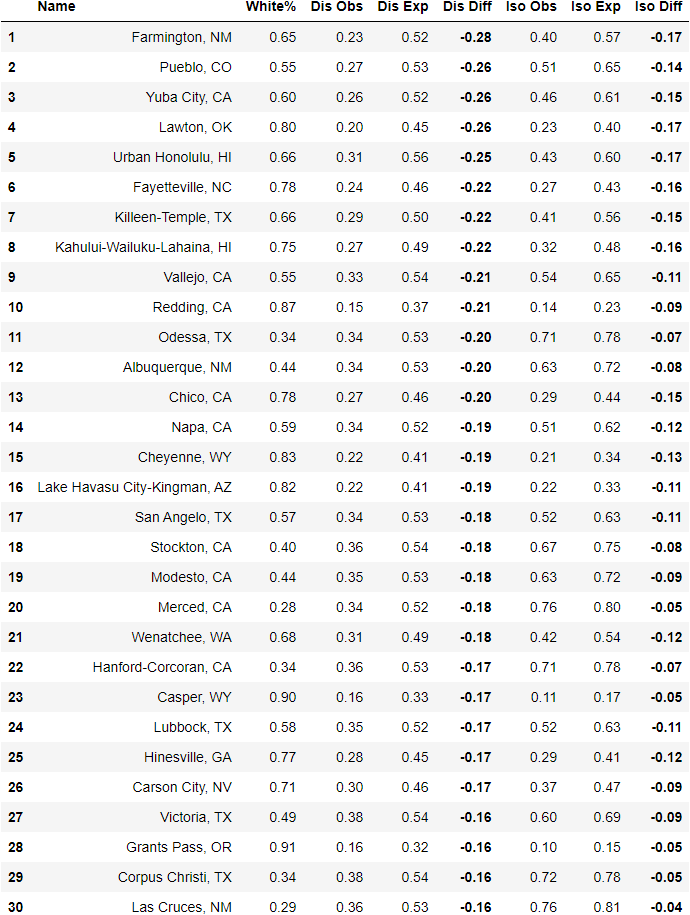}
    \textbf{Table 10}: White/Hispanic least segregated metropolitan areas in 2020 as determined by their dissimilarity index difference from 2020 White/Hispanic national expectations.
\end{figure}

\begin{figure}[hbt!]
    \includegraphics[width=6.4in]{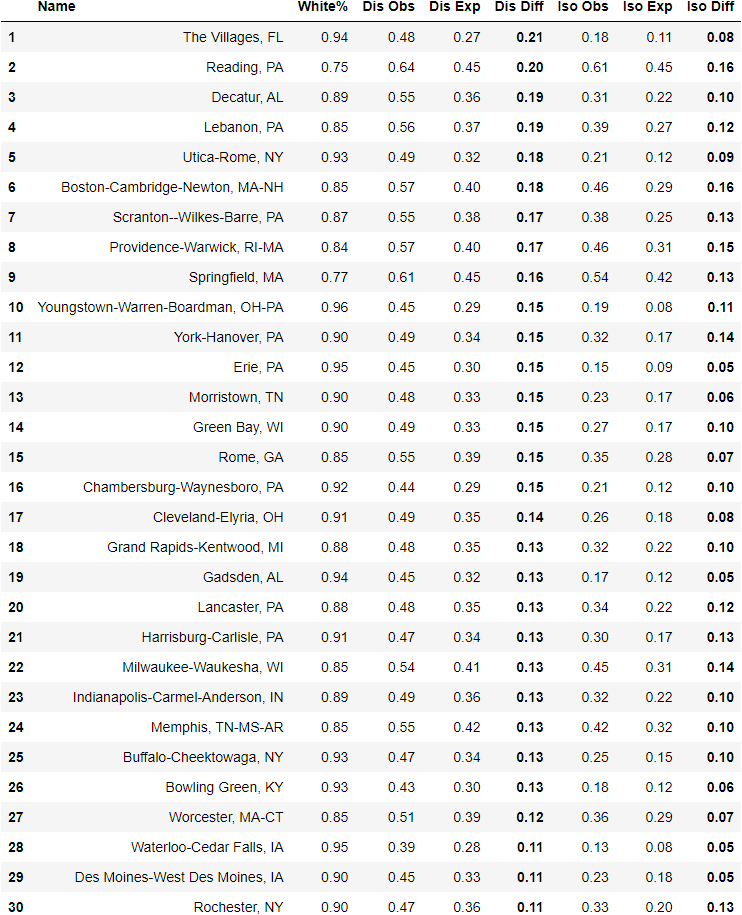}
    \textbf{Table 11}: White/Hispanic most segregated metropolitan areas in 2020 as determined by their dissimilarity index difference from 2020 White/Hispanic national expectations.
\end{figure}

\begin{figure}[hbt!]
    \includegraphics[width=6.5in,keepaspectratio]{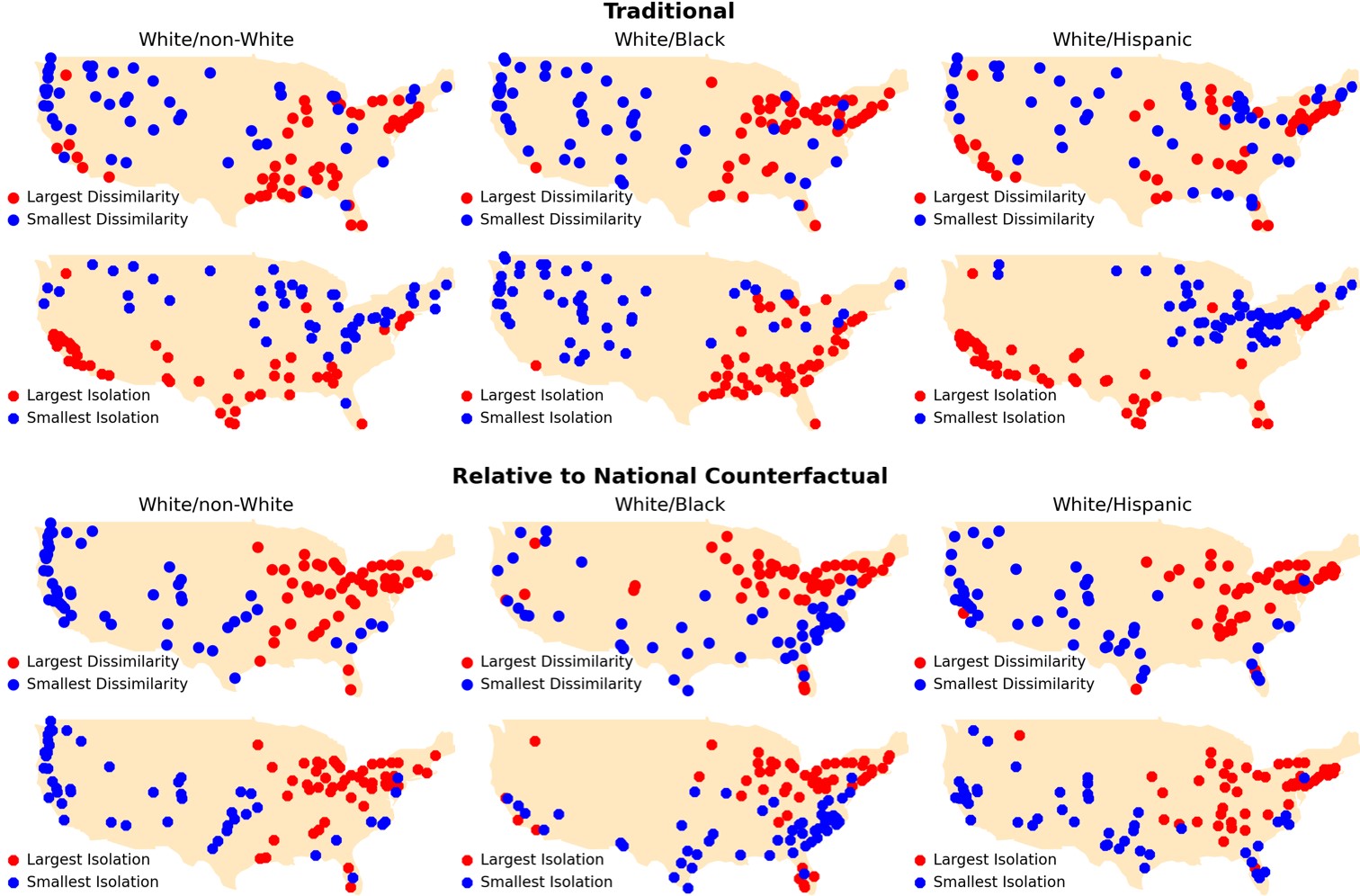}
    \caption{\textbf{Spatial clustering of extremes of least (blue) and most (red) segregated metropolitan regions:} determined from the traditional dissimilarity and isolation indices (upper panels) and counter-factual comparison to the national norm with accounting for overall regional compositions and neighborhood sizes (lower panels). Accounting for marginal differences greatly clarifies geographical trends for White/non-White, White/Black, and White/Hispanic segregation. These results demonstrate that the North-Eastern regions in the United States tend to be the most segregated and that the Southern and Western regions tend to be the least.}
    \label{fig:All least/most geographical}
\end{figure}

\end{document}